\documentclass[english]{IEEEtran}
\usepackage[T1]{fontenc}
\usepackage[latin9]{inputenc}
\usepackage{color}
\usepackage{verbatim}
\usepackage{float}
\usepackage{amsmath}
\usepackage{amsthm}
\usepackage{amssymb}
\usepackage{stmaryrd}
\usepackage{stackrel}
\usepackage{graphicx}
\usepackage{nicefrac}  
\usepackage{wasysym}
\usepackage{mathtools}
\usepackage{hyperref}
\usepackage{url}   
\usepackage{booktabs} 
\usepackage{nicefrac} 
\usepackage{subfig}
\usepackage[dvipsnames]{xcolor}
\makeatletter

\floatstyle{ruled}
\newfloat{algorithm}{tbp}{loa}
\providecommand{\algorithmname}{Algorithm}
\floatname{algorithm}{\protect\algorithmname}

\theoremstyle{plain}
\newtheorem{thm}{\protect\theoremname}
\theoremstyle{definition}
\newtheorem{defn}{\protect\definitionname}
\theoremstyle{definition}
\newtheorem{problem}{\protect\problemname}
\theoremstyle{plain}
\newtheorem{lem}{\protect\lemmaname}
\theoremstyle{definition}
\newtheorem{example}{\protect\examplename}
\theoremstyle{remark}

\theoremstyle{plain}

\usepackage{tikz}
\tikzset{>=latex}
\usetikzlibrary{fit,automata,positioning,angles,quotes}

\usepackage{cite}
\usepackage{algpseudocode}
\usepackage{setspace}
\IEEEoverridecommandlockouts

\usepackage{fancyhdr}
\makeatother

\usepackage{babel}
\providecommand{\corollaryname}{Corollary}
\providecommand{\definitionname}{Definition}
\providecommand{\examplename}{Example}
\providecommand{\lemmaname}{Lemma}
\providecommand{\problemname}{Problem}
\providecommand{\remarkname}{Remark}
\providecommand{\theoremname}{Theorem}

\usepackage{amsmath}               
  {
      \theoremstyle{plain}
      \newtheorem{assumption}{Assumption}
  }

\SetSymbolFont{stmry}{bold}{U}{stmry}{m}{n}

\def\BibTeX{{\rm B\kern-.05em{\sc i\kern-.025em b}\kern-.08em
		T\kern-.1667em\lower.7ex\hbox{E}\kern-.125emX}}
\begin{document}	
\title{Reinforcement Learning Based Temporal Logic Control with Maximum Probabilistic
Satisfaction\thanks{$^{1}$Department of Mechanical Engineering, The University of Iowa,
Iowa City, IA, USA.}\thanks{$^{2}$Department of Automation, University of Science and Technology
of China, Hefei, Anhui, China.}}
\author{Mingyu Cai$^{1}$, Shaoping Xiao$^{1}$, Baoluo
	Li$^{2}$, Zhiliang Li$^{2}$, and Zhen Kan$^{2}$}
\maketitle
\thispagestyle{fancy}

\begin{abstract}
This paper presents a model-free reinforcement learning (RL) algorithm
to synthesize a control policy that maximizes the satisfaction probability of complex tasks, which are expressed by linear temporal logic (LTL) specifications. Due to the consideration
of environment and motion uncertainties, we model the robot motion
as a probabilistic labeled Markov decision process (PL-MDP) with  unknown transition probabilities and probabilistic labeling functions. The LTL task
specification is converted to a limit deterministic generalized B\"uchi
automaton (LDGBA) with several accepting sets to maintain dense rewards
during learning. The novelty of applying LDGBA is
to construct an embedded LDGBA (E-LDGBA) by designing a synchronous
tracking-frontier function, which enables
the record of non-visited accepting sets of LDGBA at each round
of the repeated visiting pattern, to overcome the difficulties of directly applying conventional LDGBA.
With appropriate dependent reward and discount functions, rigorous
analysis shows that any method, which optimizes the expected discount
return of the RL-based approach, is guaranteed to find the optimal
policy to maximize the satisfaction probability of the LTL specifications.
A model-free RL-based motion planning strategy is developed to generate
the optimal policy in this paper. The effectiveness of the RL-based
control synthesis is demonstrated via simulation and experimental
results.

\global\long\def\Dist{\operatorname{Dist}}%
\global\long\def\Inf{\operatorname{Inf}}%
\global\long\def\Sense{\operatorname{Sense}}%
\global\long\def\Eval{\operatorname{Eval}}%
\global\long\def\Info{\operatorname{Info}}%
\global\long\def\ResetRabin{\operatorname{ResetRabin}}%
\global\long\def\Post{\operatorname{Post}}%
\global\long\def\Acc{\operatorname{Acc}}%
\end{abstract}

\section{INTRODUCTION}

Temporal logic has rich expressivity in describing complex high-level
tasks beyond traditional go-to-goal navigation for robotic systems
\cite{Baier2008,Cai2020c,Kantaros2020}. Due to a variety of uncertainties (e.g., transition
probabilities and environment uncertainties), the robot's probabilistic
motion is often modeled by a Markov decision process (MDP). Growing
research has been devoted to investigating the motion planning of
an MDP satisfying linear temporal logic (LTL) constraints. With the
assumption of full knowledge of MDP, one common objective is to maximize
the probability of accomplishing tasks \cite{Ding2014a,Guo2018,Lacerda2019,Cai2020b}.
Yet, it raises some challenges when the MDP is not fully known \textit{a
	priori}. Hence, this work focuses on motion planning that maximizes
the satisfaction probability of given tasks over an uncertain MDP.

Reinforcement learning (RL) is a widely-used approach for sequential decision-making problems \cite{Watkins1992}. When integrating with LTL specifications,
model-based RL has been employed in \cite{Sadigh2014,Fu2014,Wang2015}
to generate policies to satisfy LTL tasks by learning unknown parameters
of the MDP. However, there is a scalability issue due to the high
need of memory to store the learned models. On the other hand, model-free RL generates
policies to satisfy LTL formulas by designing appropriate accepting
rewards to optimize Q values \cite{Li2019,Gao2019,Cai2020,Hahn2019,Bozkurt2020,Hasanbeig2019a,Oura2020,Wang2020}.
In \cite{Li2019}, the robustness degree of truncated linear temporal
logic (TLTL) was used as reward to facilitate learning. The deterministic finite automaton (DFA) was applied as reward machines
in \cite{deepsynth,Icarte2018,Camacho2019}. However, only
finite horizon motion planning was considered in \cite{Li2019,deepsynth,Icarte2018,Camacho2019}.

\textbf{Related works:} This work extends previous research to
tasks over infinite horizon, where finite horizon motion planning
can be regarded as a special case of the infinite horizon setting. Along
this line of research, in \cite{Gao2019} and
\cite{Cai2020}, LTL constraints were translated to Deterministic Rabin
Automata (DRA), which may fail to find desired policies as discussed
in \cite{Hahn2019}. Instead of using DRA, limit-deterministic Buchi
automaton (LDBA) was employed in \cite{Hahn2019} and \cite{Bozkurt2020}
without considering the workspace uncertainties. Moreover, since LDBA in works \cite{Hahn2019} and \cite{Bozkurt2020} has only
one accepting set, it might lead to sparse reward issues during learning.
In \cite{Hasanbeig2019a}, limit-deterministic generalized Buchi automaton
(LDGBA) was used, and a frontier function of rewards was designed to
facilitate learning by assigning positive rewards to the accepting
sets. 
However, directly applying the LDGBA as in \cite{Hasanbeig2019a} may fail to satisfy the LTL specification when applying the deterministic policy and such a drawback was also presented in \cite{Oura2020}. Such an issue may be solved via selecting these actions based on the uniform distribution when applying the tubular RL method. However, the application of deterministic policies is crucial in practice especially for the continuous space, since many widely-applied deep RL methods adopt the actor-critic architecture e.g., as deep deterministic policy gradients (DDPG) and trust region policy optimization (TRPO) for high dimensional analysis. The works of \cite{Oura2020}
and \cite{Wang2020} overcome this issue by designing binary-valued
vectors and Boolean vectors, respectively. However, both \cite{Oura2020} and \cite{Wang2020}
cannot guarantee the maximum probability of task satisfaction.

\textbf{Contributions:} Our framework studies motion planning that maximizes the probability of satisfying pre-specified LTL tasks in stochastic systems. Considering both motion and environment uncertainties, the robotic system is modeled as a probabilistic labeled Markov decision process (PL-MDP) with unknown transition probabilities and probabilistic labels. In this work, a synchronous tracking-frontier function is designed to construct an embedded LDGBA (E-LDGBA) from convention LDGBA, which is capable of recording non-visited accepting sets and incorporating deterministic policies. We construct the embedded product MDP (EP-MDP) between E-LDGBA and PL-MDP, and  propose a new expected return by applying the reward and discount functions of \cite{Bozkurt2020}.
Rigorous analysis shows that our framework is guaranteed to find the optimal policy that maximizes the probability of satisfying LTL specifications. 

\section{PRELIMINARIES}

\subsection{Probabilistic Labeled MDP\label{subsec:Labeled-MDP}}

A PL-MDP is a tuple $\mathcal{M}=\left(S,A,p_{S},\left(s_{0},l_{0}\right),\Pi,L,p_{L}\right)$,
where $S$ is a finite state space, $A$ is a finite action space,
$p_{S}:S\times A\times S\shortrightarrow\left[0,1\right]$ is the
transition probability function, $\Pi$ is a set of atomic propositions,
and $L:S\shortrightarrow2^{\Pi}$ is a labeling function. The pair
$\left(s_{0},l_{0}\right)$ denotes an initial state $s_{0}\in S$
and an initial label $l_{0}\in L\left(s_{0}\right)$. The function
$p_{L}\left(s,l\right)$ denotes the probability of $l\subseteq L\left(s\right)$
associated with $s\in S$ satisfying $\sum_{l\in L\left(s\right)}p_{L}\left(s,l\right)=1,\forall s\in S$. For simplicity, let $A(s)$ denote the set of actions that can be taken in state $s$.
The transition probability $p_{S}$ captures the motion uncertainties
of the agent while the labeling probability $p_{L}$ captures the
environment uncertainties. It is assumed that $p_{S}$ and $p_{L}$
are not known \textit{a priori}, and the agent can only observe its
current state and the associated labels. Note that the standard MDP model can be regarded as a special case of PL-MDP with the deterministic label function.

Let $\xi$ be a deterministic action function such that $\xi:S\rightarrow A$ maps a state $s\in S$ to an action
in $A\left(s\right)$. The PL-MDP $\mathcal{M}$ evolves by taking an action $\xi_{i}$
at each stage $i$%
, and thus the control policy $\boldsymbol{\xi}=\xi_{0}\xi_{1}\ldots$
is a sequence of actions, which yields a path $\boldsymbol{s}=s_{0}s_{1}s_{2}\ldots$
over $\mathcal{M}$ with $p_{S}\left(s_{i},a_{i},s_{i+1}\right)>0$
for all $i$. If $\xi_{i}=\xi$ for all $i$, then $\boldsymbol{\xi}$
is called a stationary policy. The control policy $\boldsymbol{\xi}$
is memoryless if each $\xi_{i}$ only depends on its current state,
and $\boldsymbol{\xi}$ is called a finite memory policy if $\xi_{i}$
depends on its past states.

Let $\varLambda:S\times A\times S\shortrightarrow\mathbb{R}$ denote
a reward function. Given a discount function $\gamma:S\times A\times S\shortrightarrow\mathbb{R}$,
the expected discounted return under policy \textbf{$\boldsymbol{\xi}$} starting
from $s\in S$ is defined as $$U^{\boldsymbol{\xi}}\left(s\right)=\mathbb{E}^{\boldsymbol{\xi}}\left[\stackrel[i=0]{\infty}{\sum}\gamma^{i}\left(s_{i},a_{i},s_{i+1}\right)\cdot\varLambda\left(s_{i},a_{i},s_{i+1}\right)\left|s_{0}=s\right.\right].$$
An optimal policy $\boldsymbol{\xi}^{*}$ that maximizes the expected
return for each state $s\in S$ is defined as $$\boldsymbol{\xi}^{*}=\underset{\boldsymbol{\xi}}{\arg\max}U^{\boldsymbol{\xi}}\left(s\right).$$
The function $U^{\boldsymbol{\xi}}\left(s\right)$ is often referred
to as the value function under policy \textbf{$\boldsymbol{\xi}$}.
If the MDP is not fully known, but the state and action spaces are countably finite, tabular approaches are usually
employed \cite{Watkins1992}.

\subsection{LTL and Limit-Deterministic Generalized B\"uchi Automaton}

An LTL is built on atomic propositions, Boolean operators, and temporal
operators \cite{Baier2008}. Given an LTL that specifies the missions,
the satisfaction of the LTL can be evaluated by an LDGBA\cite{Sickert2016}.
Before defining LDGBA, we first introduce the generalized B\"uchi automaton
(GBA). 
\begin{defn}
	\label{def:GBA} A GBA is a tuple $\mathcal{A}=\left(Q,\Sigma,\delta,q_{0},F\right)$,
	where $Q$ is a finite set of states; $\Sigma=2^{\Pi}$ is a finite
	alphabet, $\delta\colon Q\times\Sigma\shortrightarrow2^{Q}$ is the
	transition function, $q_{0}\in Q$ is an initial state, and $F=\left\{ F_{1},F_{2},\ldots,F_{f}\right\} $
	is a set of accepting sets with $F_{i}\subseteq Q$, $\forall i\in\left\{ 1,\ldots f\right\} $. 
\end{defn}
Denote by $\boldsymbol{q}=q_{0}q_{1}\ldots$ a run of a GBA, where
$q_{i}\in Q$, $i=0,1,\ldots$. The run $\boldsymbol{q}$ is accepted
by the GBA, if it satisfies the generalized B\"uchi acceptance condition,
i.e., $\inf\left(\boldsymbol{q}\right)\cap F_{i}\neq\emptyset$, $\forall i\in\left\{ 1,\ldots f\right\} $,
where $\inf\left(\boldsymbol{q}\right)$ denotes the infinitely part
of $\boldsymbol{q}$. 
\begin{defn}
	\label{def:LDGBA} A GBA is an LDGBA if the transition function $\delta$
	is extended to $Q\times\left(\Sigma\cup\left\{ \epsilon\right\} \right)\shortrightarrow2^{Q}$,
	and the state set $Q$ is partitioned into a deterministic set $Q_{D}$
	and a non-deterministic set $Q_{N}$, i.e., $Q_{D}\cup Q_{N}=Q$ and
	$Q_{D}\cap Q_{N}=\emptyset$, where 
	\begin{itemize}
		\item the state transitions in $Q_{D}$ are total and restricted within
		it, i.e., $\bigl|\delta\left(q,\alpha\right)\bigr|=1$ and $\delta\left(q,\alpha\right)\subseteq Q_{D}$
		for every state $q\in Q_{D}$ and $\alpha\in\Sigma$, 
		\item the $\epsilon$-transition is not allowed in the deterministic set,
		i.e., for any $q\in Q_{D}$, $\delta\left(q,\epsilon\right)=\emptyset$,
		and 
		\item the accepting sets are only in the deterministic set, i.e., $F_{i}\subseteq Q_{D}$
		for every $F_{i}\in F$. 
	\end{itemize}
\end{defn}
In Definition \ref{def:LDGBA}, the $\epsilon$-transitions are only
defined for state transitions from $Q_{N}$ to $Q_{D}$, which do
not consume the input alphabet. To convert an LTL formula to an LDGBA,
readers are referred to Owl \cite{Kretinsky2018}.

\section{Problem Statements}

The task specification to be performed by the robot is described by
an LTL formula $\phi$ over $\Pi$. Given Task $\phi$, the PL-MDP
$\mathcal{M},$ and a policy $\boldsymbol{\xi}=\xi_{0}\xi_{1}\ldots$,
the induced path $\boldsymbol{s}_{\infty}^{\boldsymbol{\xi}}=s_{0}\ldots s_{i}s_{i+1}\ldots$
over $\mathcal{M}$ satisfies $s_{i+1}\in\left\{ s\in S\bigl|p_{S}\left(s_{i},a_{i},s\right)>0\right\} $.
Let $L\left(\boldsymbol{s}_{\infty}^{\boldsymbol{\xi}}\right)=l_{0}l_{1}\ldots$
be the sequence of labels associated with $\boldsymbol{s}_{\infty}^{\boldsymbol{\xi}}$
such that $l_{i}\in L\left(s_{i}\right)$ and 
1
 $p_{L}\left(s_{i},l_{i}\right)>0$.
Denote by $L\left(\boldsymbol{s}_{\infty}^{\boldsymbol{\xi}}\right)\models\phi$
if the induced trace $\boldsymbol{s}_{\infty}^{\boldsymbol{\xi}}$ satisfies
$\phi$. The probabilistic satisfaction under the policy $\xi$ from
an initial state $s_{0}$ can be defined as
\begin{equation}
	{\color{black}\Pr{}_{M}^{\boldsymbol{\xi}}\left(\phi\right)=\Pr{}_{M}^{\boldsymbol{\xi}}\left(L\left(\boldsymbol{s}_{\infty}^{\boldsymbol{\xi}}\right)\models\phi\left|\boldsymbol{s}_{\infty}^{\boldsymbol{\xi}}\in\boldsymbol{S}_{\infty}^{\boldsymbol{\xi}}\right|\right),}\label{eq:probabilistic-satisfaction}
\end{equation}
where $\boldsymbol{S}_{\infty}^{\boldsymbol{\xi}}$ is a set of admissible
paths from the initial state $s_{0}$ under the policy ${\boldsymbol{\xi}}$. 

\begin{assumption}\label{Assumption1}
  It is assumed that there exists at least one policy such that the induced traces satisfy task $\phi$ with non-zero probability.
\end{assumption}

Assumption 1 is a mild assumption and widely employed
in the literature (cf. \cite{Sadigh2014,Hahn2019,Bozkurt2020}), which indicates that the LTL task can be satisfied with nonzero probability. Consequently, the
following problem is considered. 
\begin{problem}
	\label{Prob1}Given an LTL-specified task $\phi$ and a PL-MDP $\mathcal{M}$
	with unknown transition probabilities (i.e., motion uncertainties)
	and an unknown probabilistic label function (i.e., workspace uncertainties),
	the objective is to find the desired policy $\boldsymbol{\xi^{*}}$ that maximizes the
	satisfaction probability, i.e., $\boldsymbol{\xi^{*}}=\underset{\boldsymbol{\xi}}{\arg\max}\Pr{}_{M}^{\boldsymbol{\xi}}\left(\phi\right)$,
	by interacting with the environment. 
\end{problem}

In order to find the desired policy in PL-MDP $\mathcal{M}$ to satisfy the user-specified LTL formula $\phi$, we can construct the standard product MDP between $\mathcal{M}$ and the LDGBA of $\phi$ as described in \cite{Baier2008,Hasanbeig2019}. Then, the problem becomes finding the policy that satisfies the accepting condition of the standard product MDP with maximum probability. However, when considering deterministic policies, directly applying LDGBA \cite{Hasanbeig2019} may fail to satisfy the LTL specifications, because there do not exist deterministic policies to select several actions with the same optimal expected return at one state. To illustrate this issue, Example \ref{example} is provided.

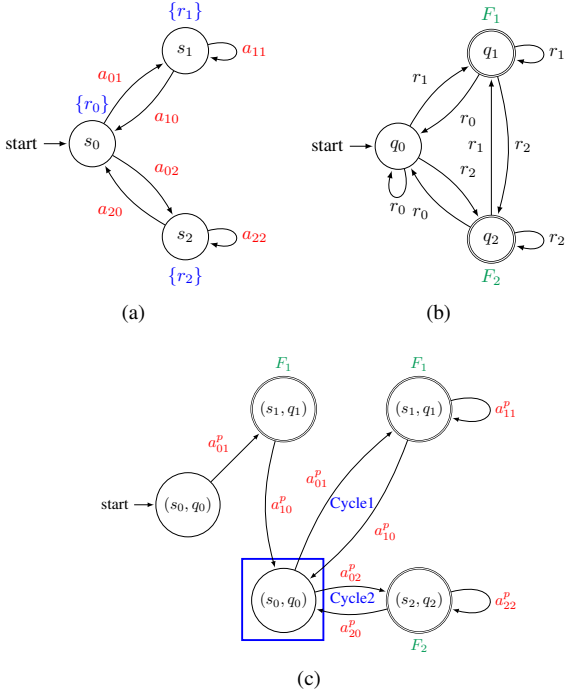
\begin{figure}[!t]\centering
	\subfloat[]{{
	    \scalebox{.7}{
	    	\begin{tikzpicture}[shorten >=1pt,node distance=2.5cm,on grid,auto] 
			\node[] (s_0)   {}; 
			\node[state, initial] (s_0) [above right=of s_0, label=above:$\textcolor{blue}{\{r_0\}}$]  {$s_0$};
			\node[state] (s_1) [above right=of s_0, label=above:$\textcolor{blue}{\{r_1\}}$]  {$s_1$};
			\node[state] (s_2) [below right=of s_0, label=below:$\textcolor{blue}{\{r_2\}}$]  {$s_2$};
			\path[->] 
			(s_0) edge [bend left=15] node {$\textcolor{red}{a_{01}}$} (s_1)
			(s_1) edge [bend left=15] node {$\textcolor{red}{a_{10}}$} (s_0)
			(s_1) edge [loop right] node {$\textcolor{red}{a_{11}}$} (s_1)
			(s_2) edge [loop right] node {$\textcolor{red}{a_{22}}$} (s_2)
			(s_0) edge [bend left=15] node {$\textcolor{red}{a_{02}}$} (s_2)
			(s_2) edge [bend left=15] node {$\textcolor{red}{a_{20}}$} (s_0);
			\end{tikzpicture}

			}

			}}
	\subfloat[]{{
		\scalebox{.7}{
			\begin{tikzpicture}[shorten >=1pt,node distance=2.5cm,on grid,auto] 
			\node[state,initial] (q_0)   {$q_0$}; 
			\node[state,accepting] (q_1) [above right=of q_0, label=above:$\textcolor{ForestGreen}{F_1}$]  {$q_1$};
			\node[state,accepting, label=below:$\textcolor{ForestGreen}{F_2}$] (q_2) [below right=of q_0]  {$q_2$};
			\path[->] 
			(q_0) edge [bend left=15] node {$r_1$} (q_1)
			(q_1) edge [bend left=15] node {$r_0$} (q_0)
			(q_0) edge [loop below] node {$r_0$} (q_0)
			(q_1) edge [loop right] node {$r_1$} (q_1)
			(q_2) edge [loop right] node {$r_2$} (q_2)
			(q_0) edge [bend left=15] node {$r_2$} (q_2)
			(q_2) edge [bend left=15] node {$r_0$} (q_0)
			(q_1) edge [bend left=15] node {$r_2$} (q_2)
			(q_2) edge node {$r_1$} (q_1);
			\end{tikzpicture}
			}
			}}
	$\qquad$
	\subfloat[]{{
		\scalebox{.6}{
			\begin{tikzpicture}[shorten >=1pt,node distance=3cm,on grid,auto] 
			\node[state,initial] (s_0)   {$(s_0,q_0)$}; 
			\node[state,accepting] (s_1) [above right=of s_0, label=above:$\textcolor{ForestGreen}{F_1}$]  {$(s_1,q_1)$};
			\node[state] (s_2) [below right=of s_0]  {$(s_0,q_0)$};
			\draw[blue, very thick] (1.2,-3) rectangle (3,-1.2);
			\path (3.65,0) node(x) {\textcolor{blue}{Cycle1}};
			\path (3.65,-2.1) node(x) {\textcolor{blue}{Cycle2}};
			\node[state,accepting] (s_3) [right=of s_1, label=above:$\textcolor{ForestGreen}{F_1}$]  {$(s_1,q_1)$};
			\node[state,accepting] (s_4) [right=of s_2, label=below:$\textcolor{ForestGreen}{F_2}$]  {$(s_2,q_2)$};
			\path[->]
			(s_0) edge node {$\textcolor{red}{a_{01}^p}$} (s_1)
			(s_1) edge [bend right=15] node {$\textcolor{red}{a_{10}^p}$} (s_2)
			(s_2) edge [bend left=15] node {$\textcolor{red}{a_{01}^p}$} (s_3)
			(s_3) edge [bend left=15] node {$\textcolor{red}{a_{10}^p}$} (s_2)
			(s_3) edge [loop right] node {$\textcolor{red}{a_{11}^p}$} (s_3)
			(s_2) edge [bend left=15] node {$\textcolor{red}{a_{02}^p}$} (s_4)
			(s_4) edge [bend left=15] node {$\textcolor{red}{a_{20}^p}$} (s_2)
			(s_4) edge [loop right] node {$\textcolor{red}{a_{22}^p}$} (s_4)
			;
			\end{tikzpicture}
			}
		}}
		\caption{\label{fig:automaton} (a) PL-MDP model with the deterministic label function. (b) LDGBA of LTL formula
		$\varphi_{e}$. (c) The standard product MDP.}
	\end{figure}
\begin{example}
\label{example}
Here  is  an  example  to  demonstrate  why the LDGBA does not work in some cases for deterministic policies. Fig. \ref{fig:automaton} (a) shows a special case of the PL-MDP model that deterministically labels each state, in which there are three states $s_0$, $s_1$ and $s_2$ associated with three labels $r_0$, $r_1$, and $r_2$, respectively. The initial state of PL-MDP is $s_0$. The LTL specification of the PL-MDP is $\varphi_{e}=\left(\oblong\lozenge r_1\right)\land\left(\oblong\lozenge r_2\right)$, which requires the agent starting from $s_0$ labeled with $r_0$ to repetitively visit the states with labels $r_1$ and $r_2$. Fig. \ref{fig:automaton} (b) shows the corresponding LDGBA of $\varphi_{e}$ with two accepting sets $F=\left\{ \left\{ q_{1}\right\} ,\left\{ q_{2}\right\} \right\}$. Fig. \ref{fig:automaton} (c) illustrates the resulted standard
product MDP. By Def. \ref{def:LDGBA}, the policy that satisfies $\varphi_{e}$
should enforce the repetitive trajectories, i.e., Cycles 1 and 2
in Fig. \ref{fig:automaton} (c). However, there exists no deterministic
policy that can periodically select two actions $a_{01}^{P}$ and
$a_{02}^{P}$ at state $\left(s_{0},q_{0}\right)$ (marked with a
blue rectangle) in Fig. \ref{fig:automaton} (c). As a result, applying the
standard product MDP cannot generate a pure deterministic optimal policy to complete Task $\varphi_{e}$. 
\end{example}

Motivated by the challenge illustrated above, this paper proposes a novel automaton structure based on LDGBA to generate a deterministic policy to solve Problem \ref{Prob1}. We also verify that the designed automaton structure accepts the same languages as LDGBA, which will not influence the optimal convergence of the RL agent.

\section{Automaton Analysis\label{sec:Automaton-Analysis}}

To solve Problem \ref{Prob1}, Section \ref{Embedded-LDGBA} first
presents how the LDGBA in Definition \ref{def:LDGBA} can be extended
to an E-LDGBA, which keeps track of non-visited accepting sets and accepts the same language as the LDGBA. Section
\ref{subsec:PMDP} presents the construction of a EP-MDP between
a PL-MDP and an E-LDGBA. The benefits of incorporating E-LDGBA are
discussed in Section \ref{subsec:Property_relax}.
\subsection{E-LDGBA\label{Embedded-LDGBA}} 

Given an LDGBA $\mathcal{A}=\left(Q,\Sigma,\delta,q_{0},F\right)$,
inspired by \cite{Hasanbeig2019a}, a tracking-frontier set $T$ is
designed to keep track of non-visited accepting sets. Particularly,
$T$ is initialized as $F$, which is then updated based on 
\begin{equation}
	f_{V}\left(q,T\right)=\left\{ \begin{array}{cc}
		T\setminus F_{j}, & \text{if }q\in F_{j}\text{ and }F_{j}\in T,\\
		F\setminus F_{j}, & \text{if }\ensuremath{q\in F_{j}\text{ and }T=\emptyset},\\
		T, & \text{otherwise. }
	\end{array}\right.\label{eq:Trk-fontier}
\end{equation}
Once an accepting set $F_{j}$ is visited, it will be removed from
$T$. If $T$ becomes empty, it will be reset as $F\setminus F_{j}$. Since the acceptance
condition of LDGBA requires to infinitely visit all accepting sets,
we call it one round if all accepting sets have been visited (i.e.,
a round ends if $T$ becomes empty). If a state $q$ belongs to multiple
sets of $T$, all of these sets should be removed from $T$. Based on
(\ref{eq:Trk-fontier}), the E-LDGBA is constructed as follows.

\begin{defn}[Embedded LDGBA]
	\label{def:E-LDGBA} Given an LDGBA $\mathcal{A}=\left(Q,\Sigma,\delta,q_{0},F\right)$,
	its corresponding E-LDGBA is denoted by $\mathcal{\overline{A}}=\left(\overline{Q},\Sigma,\overline{\delta},\overline{q_{0}},\overline{F},f_{V},T\right)$
	where $T$ is initially set as $T=F$; $\overline{Q}=Q\times2^{Q}$ is the set of augmented states e.g., $\overline{q}=(q,T)$; The finite alphabet $\Sigma$ is the same as the LDGBA; The transition $\overline{\delta}\colon \overline{Q}\times\left(\Sigma\cup\left\{ \epsilon\right\} \right)\shortrightarrow2^{\overline{Q}}$
	is defined as $\overline{q'}=\overline{\delta}\left(\overline{q},\overline{\sigma}\right)$ with $\overline{\sigma}\in(\Sigma\cup\left\{ \epsilon\right\})$, e.g., $\overline{q}=(q,T)$ and $\overline{q'}=(q',T)$, and it satisfies
	two conditions: 1) $q'=\delta\left(q,\overline{\sigma}\right)$, and 2) $T$ is synchronously updated as $T=f_{V}\left(q',T\right)$ after transition $\overline{q'}=\overline{\delta}\left(\overline{q},\alpha\right)$; $\overline{F}=\left\{ \overline{F_{1}},\overline{F_{2}}\ldots \overline{F_{f}}\right\} $
where $\overline{F_{j}}=\left\{ \left(q, T\right)\in \overline{Q}\bigl|q\in F_{j}\land F_{j} \subseteq T\right\} $,
$j=1,\ldots f$, is a set of accepting states. 
\end{defn}
\begin{algorithm}
	\caption{\label{Alg:LDGBA} \textcolor{black}{Procedure of E-LDGBA}}
	
	\scriptsize
	
	\singlespacing
	
	\begin{algorithmic}[1]
		
		\Procedure {Input: } {LDGBA $\mathcal{A}$, $f_{V},T$ and length
			$L$}
		
		{Output: } { A valid run $\boldsymbol{\overline{q}}$ with
			length $L$ in $\mathcal{\overline{A}}_{\phi}$ }
		\State set $T=F$ and $count=1$
		
		\State set $\overline{q}_{cur}=(q_{0},T)$ and $\boldsymbol{\overline{q}}=\left(q_{cur}\right)$
		
		\State set $q_{cur}=q$ 
		
		\While { $count\leq L$ }
		
		\State $q_{next}=\delta(q_{cur},\alpha)$

		\State $\overline{q}_{cur}\shortleftarrow (q_{next},T)$
		
		\State check if $\overline{q}_{cur}$ is an accepting state
		
		\State$T=f_{V}\left(q_{next},T\right)$
		
	    \State add state $\overline{q}_{cur}$ to $\boldsymbol{\overline{q}}$
		
		\State $count++$ and $q_{cur}\shortleftarrow q_{next}$
		
		\EndWhile
		
		\EndProcedure
		
	\end{algorithmic}
\end{algorithm}

 In Definition \ref{def:E-LDGBA}, we abuse the tuple structure since the frontier set $T$ is synchronously updated after each transition, and each state of E-LDGBA is augmented with the tracking-frontier set $T$ at every time-step via one-hot encoding. The accepting state is determined based on the current automaton state, and frontier set $T$ that is synchronously updated after each transition. 
 Such property
is the innovation of E-LDGBA, which encourages all accepting sets to be
visited in each round. 
In the following analysis, we will use $\mathcal{\overline{A}}_{\phi}$ and $\mathcal{A}_{\phi}$
to denote the E-LDGBA and LDGBA, respectively, corresponding to an LTL formula $\phi$. Algorithm \ref{Alg:LDGBA} shows the procedure of obtaining a valid
run $\boldsymbol{\overline{q}}$ over an E-LDBGA $\mathcal{\overline{A}}_{\phi}$. 

Given $\mathcal{\overline{A}}_{\phi}$ and $\mathcal{A}_{\phi}$ for the same LTL formula, the E-LDGBA $\mathcal{\overline{A}}_{\phi}$ keeps track of unvisited accepting sets of  $\mathcal{A}_{\phi}$ by incorporating $f_{V}$ and $T$. The $T$ will be reset when all the accepting sets of $\mathcal{A}_{\phi}$ have been visited. Let $\mathcal{L}(\mathcal{A}_{\phi})\subseteq \Sigma^{\omega}$ and $\mathcal{L}(\mathcal{\overline{A}}_{\phi})\subseteq \Sigma^{\omega}$ be the accepted language of the  $\mathcal{A}_{\phi}$ and $\mathcal{\overline{A}}_{\phi}$ automaton, respectively, with the same alphabet $\Sigma$. Based on \cite{Baier2008}, $\mathcal{L}(\mathcal{A}_{\phi})\subseteq \Sigma^{\omega}$ is the set of all infinite words accepted by $\mathcal{A}_{\phi}$ that satisfy LTL formula $\phi$.

\begin{lem}
\label{lem:language}
For any LTL formula $\phi$, we can construct LDGBA $\mathcal{A}_{\phi}=\left(Q,\Sigma,\delta,q_{0},F\right)$ and E-LDGBA $\mathcal{\overline{A}}_{\phi}=\left(\overline{Q},\Sigma,\overline{\delta},\overline{q_{0}},\overline{F},f_{V},T\right)$. Then it holds that
\begin{equation}
\mathcal{L}(\mathcal{\overline{A}}_{\phi})=\mathcal{L}(\mathcal{A}_{\phi}).
\end{equation}
\end{lem}
\begin{IEEEproof}
We prove (\ref{lem:language}) by showing that $\mathcal{L}(\mathcal{\overline{A}}_{\phi})\supseteq\mathcal{L}(\mathcal{A}_{\phi})$ and $\mathcal{L}(\mathcal{\overline{A}}_{\phi})\subseteq\mathcal{L}(\mathcal{A}_{\phi})$.

\textbf{Case 1:}  $\mathcal{L}(\mathcal{\overline{A}}_{\phi})\supseteq\mathcal{L}(\mathcal{A}_{\phi})$: For any accepting language $\boldsymbol{\omega}=\alpha_{0}\alpha_{1}\ldots\in\mathcal{L}(\mathcal{A}_{\phi})$, there exists a corresponding run $\boldsymbol{r}=q_{0}\alpha_{0}q_{1}\alpha_{1}\ldots$ of $\mathcal{A}_{\phi}$ s.t.
\begin{equation}
\inf\left(\boldsymbol{r}\right)\cap F_{i}\neq\emptyset, \forall i\in\left\{ 1,\ldots f\right\}. \label{eq:Lemma_case1}
\end{equation}
For the run $\boldsymbol{r}$, we can construct a sequence $\boldsymbol{\overline{r}}=\overline{q}_{0}\alpha_{0}\overline{q}_{1}\alpha_{1}\ldots$ by add each state $q$ with the set $T$, which is synchronously updated via (\ref{eq:Trk-fontier}) after each transition. It can be verified that such a run $\boldsymbol{\overline{r}}$ is a valid run of $\mathcal{\overline{A}}_{\phi}$ based on Def. \ref{def:E-LDGBA}. According to (\ref{eq:Lemma_case1}), since the tracking-frontier set $T$ will be reset once all accepting sets have been visited, it holds $\inf\left(\boldsymbol{\overline{r}}\right)\cap \overline{F_{i}}\neq\emptyset, \forall i\in\left\{ 1,\ldots f\right\}$ s.t. $\boldsymbol{\omega}\in\mathcal{L}(\mathcal{\overline{A}}_{\phi})$.  

\textbf{Case 2:} $\mathcal{L}(\mathcal{\overline{A}}_{\phi})\subseteq\mathcal{L}(\mathcal{A}_{\phi})$: Similarly, for any accepting language $\boldsymbol{\overline{\omega}}=\overline{\alpha}_{0}\overline{\alpha}_{1}\ldots\in\mathcal{L}(\mathcal{\overline{A}}_{\phi})$, there exists a corresponding run $\boldsymbol{\overline{r}}=\overline{q}_{0}\overline{\alpha}_{0}\overline{q}_{1}\overline{\alpha}_{1}\ldots$ of $\mathcal{\overline{A}}_{\phi}$ s.t.
\begin{equation}
\inf\left(\boldsymbol{\overline{r}}\right)\cap \overline{F_{i}}\neq\emptyset, \forall i\in\left\{ 1,\ldots f\right\}. \label{eq:Lemma_case2}
\end{equation}
For the run $\boldsymbol{\overline{r}}$, we can construct a sequence $\boldsymbol{r}=q_{0}\overline{\alpha}_{0}q_{1}\overline{\alpha}_{1}\ldots$ by projecting each state $\overline{q}=(q,T)$ into $q$. It can be simply verified that such a run $\boldsymbol{r}$ is a valid run of $\mathcal{A}_{\phi}$ based on Def. \ref{def:E-LDGBA}. According to (\ref{eq:Lemma_case2}), it holds $\inf\left(\boldsymbol{r}\right)\cap F_{i}\neq\emptyset, \forall i\in\left\{ 1,\ldots f\right\}$ s.t. $\boldsymbol{\overline{\omega}}\in\mathcal{L}(\mathcal{A}_{\phi})$. 
\end{IEEEproof}

Lemma \ref{lem:language} indicates that both E-LDGBA and LDGBA accept the same language. Consequently, E-LDGBA can also be applied to verify the satisfaction of LTL specifications, and incorporating E-LDGBA into RL based model checking will not affect the convergence of optimality.

\subsection{Embedded Product MDP\label{subsec:PMDP}}
\begin{defn}
	\label{def:P-MDP} Given a PL-MDP $\mathcal{M}$ and an E-LDGBA $\mathcal{\overline{A}}_{\phi}$,
	the embedded product MDP (EP-MDP) is defined as $\mathcal{P}=\mathcal{M}\times\mathcal{\overline{A}}_{\phi}=\left(X,U^{\mathcal{P}},p^{\mathcal{P}},x_{0},F^{\mathcal{P}}, T, f_{V}\right)$,
	where $X=S\times2^{\Pi}\times \overline{Q}$ is the set of labeled states, i.e.,
	$x=\left(s,l,q,T\right)\in X$ with $l\in L\left(s\right)$ satisfying
	$p_{L}\left(s,l\right)>0$; $U^{\mathcal{P}}=A\cup\left\{ \epsilon\right\} $
	is the set of actions, where the $\epsilon$-transitions are only allowed
	for transitions from $Q_{N}$ to $Q_{D}$; $x_{0}=\left(s_{0},l_{0},\overline{q_{0}}\right)$
	is the initial state; $F^{\mathcal{P}}=\left\{ F_{1}^{\mathcal{P}},F_{2}^{\mathcal{P}}\ldots F_{f}^{\mathcal{P}}\right\} $
	where $F_{j}^{\mathcal{P}}=\left\{ \left(s,l,\overline{q}\right)\in X\bigl|\overline{q}\in \overline{F_{j}}\right\} $,
	$j=1,\ldots f$, is the set of accepting states; $p^{\mathcal{P}}:X\times U^{\mathcal{P}}\times X\shortrightarrow\left[0,1\right]$
	is transition probability defined as: 1) $p^{\mathcal{P}}\left(x,u^{\mathcal{P}},x'\right)=p_{L}\left(s',l'\right)\cdotp p_{S}\left(s,a,s^{\prime}\right)$
	if $\overline{\delta}\left(q,l\right)=q^{\prime}$ and $u^{\mathcal{P}}=a\in A\left(s\right)$;
	2) $p^{\mathcal{P}}\left(x,u^{\mathcal{P}},x'\right)=1$ if $\ensuremath{u^{\mathcal{P}}\in\left\{ \epsilon\right\} }$,
	$q'\in\overline{\delta}\left(q,\epsilon\right)$, and $\left(s',l'\right)=\left(s,l\right)$;
	and 3) $p^{\mathcal{P}}\left(x,u^{\mathcal{P}},x'\right)=0$ otherwise. After completing each transition $q'={\delta}\left(q,\alpha\right)$ based on $\overline{\delta}$,
$T$ is synchronously updated as ${\left(T\right)=f_{V}\left(q',T\right)}$
by (\ref{eq:Trk-fontier}). 
\end{defn}
The EP-MDP $\mathcal{P}$ captures the intersections between
all feasible paths over $\mathcal{M}$ and all words accepted to $\overline{\mathcal{A}}_{\phi}$,
facilitating the identification of admissible agent motions that satisfy
task $\phi$. Let $\boldsymbol{\pi}$ denote a policy over $\mathcal{P}$
and denote by $\boldsymbol{x}_{\infty}^{\boldsymbol{\pi}}=x_{0}\ldots x_{i}x_{i+1}\ldots$
the infinite path generated by $\boldsymbol{\pi}$. A path $\boldsymbol{x}_{\infty}^{\boldsymbol{\pi}}$
is accepted if $\inf\left(\boldsymbol{x}_{\infty}^{\boldsymbol{\pi}}\right)\cap F_{i}^{\mathcal{P}}\neq\emptyset$
, $\forall i\in\left\{ 1,\ldots f\right\} $. The accepting run $\boldsymbol{x}_{\infty}^{\pi}$
can yield a policy $\boldsymbol{\xi}$ in $\mathcal{M}$ that satisfies
$\phi$. We denote $\Pr^{\mathbf{\boldsymbol{\pi}}}\left[x\models\Acc_{p}\right]$
as the probability of satisfying the acceptance of $\mathcal{P}$
under policy $\boldsymbol{\pi}$, and denote $\Pr_{max}\left[x\models\Acc_{p}\right]=\underset{\boldsymbol{\pi}}{\max}\Pr_{M}^{\boldsymbol{\pi}}\left(\Acc_{p}\right)$.

Consider a sub-EP-MDP $\mathcal{P}'_{\left(X',U'\right)}$, where
$X'\subseteq X$ and $U'\subseteq U^{\mathcal{P}}$. If $\mathcal{P}'_{\left(X',U'\right)}$
is a maximum end component (MEC) of $\mathcal{P}$ and $X'\cap F_{i}^{\mathcal{P}}\neq\emptyset$,$\forall i\in\left\{ 1,\ldots f\right\} $,
then $\mathcal{P}'_{\left(X',U'\right)}$ is called an accepting maximum
end component (AMEC) of $\mathcal{P}$. Once a path enters an AMEC,
the subsequent path will stay within it by taking restricted actions
from $U'$. There exist policies such that any state $x\in X'$ can
be visited infinitely often. As a result, satisfying the task $\phi$
is equivalent to reaching an AMEC. Moreover, an MEC that does not
contain any accepting set is called a rejecting maximum end component (RMEC) and an MEC with only partial accepting sets is called a neutral
maximum end component (NMEC) \cite{Baier2008}.
Consequently, problem \ref{Prob1} can be reformulated as follows. 
\begin{problem}
	\label{Prob:2} Given a user-specified LTL task $\phi$ and the PL-MDP
	with unknown transition probabilities (i.e., motion uncertainties)
	and unknown labeling probabilities (i.e., environment uncertainties),
	the goal is to find a policy $\boldsymbol{\pi}^{*}$ satisfying the
	acceptance condition of $\mathcal{\mathcal{P}}$ with a maximum probability,
	i.e., $\Pr^{\boldsymbol{\pi}^{*}}\left[x\models\Acc_{p}\right]=\Pr_{max}\left[x\models\Acc_{p}\right]$. 
\end{problem}

\subsection{Properties of EP-MDP\label{subsec:Property_relax}}

\begin{defn}\label{def:induced_markov_chain}
Let $MC_{\mathcal{\mathcal{P}}}^{\boldsymbol{\pi}}$ denote the Markov
chain induced by a policy $\boldsymbol{\pi}$ on $\mathcal{\mathcal{\mathcal{P}}}$,
whose states can be represented by a disjoint union of a transient
class $\ensuremath{\mathcal{T}_{\boldsymbol{\pi}}}$ and $n_R$ closed
irreducible recurrent classes $\ensuremath{\mathcal{R}_{\boldsymbol{\pi}}^{j}}$,
$j\in\left\{ 1,\ldots,n_{R}\right\} $ \cite{Kearns2002}.
\end{defn}
\begin{lem}
	\label{lemma:accepting set}Given an EP-MDP $\mathcal{\mathcal{\mathcal{P}}=M}\times\mathcal{\overline{A}}_{\phi}$
	, the recurrent class $R_{\pi}^{j}$ of $MC_{\mathcal{\mathcal{P}}}^{\boldsymbol{\pi}}$,
	$\forall j\in\left\{ 1,\ldots,n\right\} $, induced by $\pi$ satisfies
	one of the following conditions: 
	\begin{enumerate}
		\item $\ensuremath{R_{\boldsymbol{\pi}}^{j}}\cap F_{i}^{\mathcal{\mathcal{P}}}\neq\emptyset,\forall i\in\left\{ 1,\ldots f\right\} $,
		or 
		\item $R_{\boldsymbol{\pi}}^{j}\cap F_{i}^{\mathcal{\mathcal{P}}}=\emptyset,\forall i\in\left\{ 1,\ldots f\right\} $. 
	\end{enumerate}
\end{lem}
\begin{IEEEproof}
	The strategy of the following proof is based on contradiction. Assume
	there exists a policy such that $\ensuremath{\ensuremath{R_{\boldsymbol{\pi}}^{j}}}\cap F_{k}^{\mathcal{\mathcal{P}}}\neq\emptyset$,
	$\forall k\in K$, where $K$ is a subset of $2^{\left\{ 1,\ldots f\right\} }\setminus\left\{ \left\{ 1,\ldots f\right\} ,\emptyset\right\} $.
	As discussed in \cite{Durrett1999}, for each state in recurrent class,
	it holds that $\stackrel[n=0]{\infty}{\sum}p^{n}\left(x,x\right)=\infty$,
	where $x\in\ensuremath{\ensuremath{R_{\pi}^{j}}}\cap F_{k}^{\mathcal{\mathcal{P}}}$
	and $p^{n}\left(x,x\right)$ denotes the probability of returning
	from a transient state $x$ to itself in $n$ steps. This means that
	each state in the recurrent class occurs infinitely often. However,
	based on the embedded tracking-frontier function of E-LDGBA in Def.
	\ref{def:E-LDGBA}, the tracking set $T$ will not be reset until
	all accepting sets have been visited. As a result, $x_{k}=\left(s,q_{k}\right)\in\ensuremath{\ensuremath{R_{\boldsymbol{\pi}}^{j}}}\cap F_{k}^{\mathcal{\mathcal{P}}}$
	with $s\in S$ will not occur infinitely, which contradicts the property
	$\stackrel[n=0]{\infty}{\sum}p^{n}\left(x_{k},x_{k}\right)=\infty$. 
\end{IEEEproof}
Lemma \ref{lemma:accepting set} indicates that, for any policy, all
accepting sets will be placed either in the transient class or in
one of the recurrent classes.

\section{Learning-based Control Synthesis\label{sec:Solution}}

In this section, we discuss a design of reward and discount functions, and present rigorous analysis to show how such a design can guide the agent over the EP-MDP to find an optimal policy whose traces satisfy the LTL task with a maximum probability. Reinforcement learning is leveraged to identify policies for Problem \ref{Prob:2}.

\subsection{Reward Design\label{subsec:RL-reward}}

Let $F_{U}^{\mathcal{\mathcal{P}}}$ denote the union of accepting states, i.e.,
$F_{U}^{\mathcal{\mathcal{P}}}=\left\{ x\in X \bigl| x\in F_{i}^{\mathcal{\mathcal{P}}},\forall i\in\left\{ 1,\ldots f\right\}\right\} $. 
Inspired by \cite{Bozkurt2020}, we propose a reward function as:
\begin{equation}
R\left(x\right)=\left\{ \begin{array}{cc}
1-r_{F}, & \text{if }x\in F_{U}^{\mathcal{\mathcal{P}}},\\
0, & \text{otherwise,}
\end{array}\right.\label{eq:reward_function}
\end{equation}
and a discount function as 
\begin{equation}
\gamma\left(x\right)=\left\{ \begin{array}{cc}
r_{F}, & \text{if }x\in F_{U}^{\mathcal{\mathcal{P}}},\\
\gamma_{F}, & \text{otherwise,}
\end{array}\right.\label{eq:discount_function}
\end{equation}
where $r_{F}\left(\gamma_{F}\right)$ is a function of
$\gamma_{F}$ satisfying $\underset{\gamma_{F}\shortrightarrow1^{-}}{\lim}r_{F}\left(\gamma_{F}\right)=1$
and $\underset{\gamma_{F}\shortrightarrow1^{-}}{\lim}\frac{1-\gamma_{F}}{1-r_{F}\left(\gamma_{F}\right)}=0$.

Given a path $\boldsymbol{x}_{t}=x_{t}x_{t+1}\ldots$ starting from
$x_{t}$, the return is denoted by
\begin{equation}
\mathcal{D}\left(\boldsymbol{x}_{t}\right)\coloneqq\stackrel[i=0]{\infty}{\sum}\left(\stackrel[j=0]{i-1}{\prod}\gamma\left(\boldsymbol{x}_{t}\left[t+i\right]\right)\cdotp R\left(\boldsymbol{x}_{t}\left[t+j\right]\right)\right)\label{eq:DisctRetrn}
\end{equation}
where it holds $\stackrel[j=0]{-1}{\prod}\coloneqq1$, and $\boldsymbol{x}_{t}\left[t+i\right]$
denotes the $\left(i+1\right)$th state in $\boldsymbol{x}_{t}$.
Based on (\ref{eq:DisctRetrn}), the expected return of any state
$x\in X$ under policy $\pi$ can be defined as 
\begin{equation}
U^{\boldsymbol{\pi}}\left(x\right)=\mathbb{E}^{\boldsymbol{\pi}}\left[\mathcal{D}\left(\boldsymbol{x}_{t}\right)\left|\boldsymbol{x}_{t}\left[t\right]=x\right|\right].\label{eq:ExpRetrn}
\end{equation}

A bottom strongly connected component (BSCC) of the Markov chain $MC_{\mathcal{\mathcal{P}}}^{\boldsymbol{\pi}}$ (Definition \ref{def:induced_markov_chain}) is a strongly connected component with no outgoing transitions.

\begin{lem}
\label{lemma:1} For any path $\boldsymbol{x}_{t}$ and $\mathcal{D}\left(\boldsymbol{x}_{t}\right)$
in (\ref{eq:DisctRetrn}), it holds that $0\leq\gamma_{F}\cdot\mathcal{D}\left(\boldsymbol{x}_{t}\left[t+1:\right]\right)\leq\mathcal{D}\left(\boldsymbol{x}_{t}\right)\leq1-r_{F}+r_{F}\cdot\mathcal{D}\left(\boldsymbol{x}_{t}\left[t+1:\right]\right)\leq1$,
where $\boldsymbol{x}_{t}\left[t+1:\right]$ denotes the suffix of
$\boldsymbol{x}_{t}$ starting from $x_{t+1}$. Let $BSCC\left(MC_{\mathcal{\mathcal{P}}}^{\boldsymbol{\pi}}\right)$
denote the set of all BSCCs of an induced Markov chain $MC_{\mathcal{\mathcal{P}}}^{\boldsymbol{\pi}}$ and let $X_{\mathcal{\mathcal{\mathcal{P}}}}^{\boldsymbol{\pi}}$ denotes the set of accepting states that belongs to a BSCC of $MC_{\mathcal{\mathcal{P}}}^{\boldsymbol{\pi}}$ s.t. $X_{\mathcal{\mathcal{\mathcal{P}}}}^{\boldsymbol{\pi}}\coloneqq\left\{ x\in X \bigl| x\in F_{U}^{\mathcal{\mathcal{P}}} \cap BSCC\left(MC_{\mathcal{\mathcal{P}}}^{\boldsymbol{\pi}}\right) \right\} $. Then, for any states $x\in X_{\mathcal{\mathcal{\mathcal{P}}}}^{\boldsymbol{\pi}}$,
it holds that$\underset{\gamma_{F}\shortrightarrow1^{-}}{\lim}U^{\boldsymbol{\pi}}\left(x\right)=1$.

\end{lem}
The proof of Lemma \ref{lemma:1} is omitted since it is a straightforward
extension of Lemma 2 and Lemma 3 in \cite{Bozkurt2020}, by replacing
LDBA with LDGBA. Since we apply the LDGBA with several accepting sets which might result in more complicated situations, e.g., AMEC, NMEC and RMEC, we can not obtain the same results as in \cite{Bozkurt2020}. We then establish the following theorem which
is one of the main contributions.

\begin{thm}
	\label{lemma:probability} Given the EP-MDP $\mathcal{\mathcal{\mathcal{P}}=M}\times\mathcal{\overline{A}}_{\phi}$,
	for any state $x\in X$, the expected return under any policy $\pi$
	satisfies 
	\begin{equation}
		\exists i\in\left\{ 1,\ldots f\right\} ,\underset{\gamma_{F}\shortrightarrow1^{-}}{\lim}U^{\boldsymbol{\pi}}\left(x\right)=\Pr{}^{\boldsymbol{\pi}}\left[\diamondsuit F_{i}^{\mathcal{\mathcal{P}}}\right],\label{eq:reachability}
	\end{equation}
	where $\Pr^{\boldsymbol{\pi}}\left[\diamondsuit F_{i}^{\mathcal{\mathcal{P}}}\right]$
	is the probability that the paths starting from state $x$ will eventually
	intersect any one $F_{i}^{\mathcal{\mathcal{P}}}$ of $F^{\mathcal{P}}$. 
\end{thm}
\begin{IEEEproof}
Based on whether or not the path $\boldsymbol{x}_{t}$ intersects
with accepting states of $F_{i}^{\mathcal{\mathcal{P}}}$, the expected
return in (\ref{eq:ExpRetrn}) can be rewritten as 
\begin{equation}
\begin{aligned}U^{\boldsymbol{\pi}}\left(x\right)= & \mathbb{E}^{\boldsymbol{\pi}}\left[\mathcal{D}\left(\boldsymbol{x}_{t}\right)\left|\boldsymbol{x}_{t}\models\diamondsuit F_{i}^{\mathcal{\mathcal{P}}}\right.\right]\cdot\Pr{}^{\boldsymbol{\pi}}\left[x\mid=\diamondsuit F_{i}^{\mathcal{\mathcal{P}}}\right]\\
 & +\mathbb{E}^{\boldsymbol{\pi}}\left[\mathcal{D}\left(\boldsymbol{x}_{t}\right)\left|\boldsymbol{x}_{t}\neq\diamondsuit F_{i}^{\mathcal{\mathcal{P}}}\right.\right]\cdot\Pr{}^{\boldsymbol{\pi}}\left[x\mid\neq\diamondsuit F_{i}^{\mathcal{\mathcal{P}}}\right]
\end{aligned}
\label{eq:proof 1}
\end{equation}
where $\Pr^{\boldsymbol{\pi}}\left[x\models\diamondsuit F_{i}^{\mathcal{\mathcal{P}}}\right]$
and $\Pr^{\boldsymbol{\pi}}\left[x\mid\neq\diamondsuit F_{i}^{\mathcal{\mathcal{P}}}\right]$
represent the probability of eventually reaching and not reaching $F_{i}^{\mathcal{\mathcal{P}}}$
under policy $\pi$ starting from state $x$, respectively.

To find the lower bound of $U^{\boldsymbol{\pi}}\left(x\right)$,
for any $\boldsymbol{x}_{t}$ with $\boldsymbol{x}_{t}\left[t\right]=x$,
let $t+N_{t}$ be the index that $\boldsymbol{x}_{t}$ first intersects
a state in $X_{\mathcal{\mathcal{\mathcal{P}}}}^{\boldsymbol{\pi}}$,
i.e., $N_{t}=\min\left[i\bigl|\boldsymbol{x}_{t}\left[t+i\right]\in X_{\mathcal{\mathcal{\mathcal{P}}}}^{\boldsymbol{\pi}}\right]$.
The following holds
\begin{equation}
\begin{aligned} & \mathbb{E}^{\boldsymbol{\pi}}\left[\mathcal{D}\left(\boldsymbol{x}_{t}\right)\left|\boldsymbol{x}_{t}\models\diamondsuit F_{i}^{\mathcal{\mathcal{P}}}\right.\right]\\
 & \overset{_{^{\left(1\right)}}}{\geq}\mathbb{E}^{\boldsymbol{\pi}}\left[\mathcal{D}\left(\boldsymbol{x}_{t}\right)\left|\boldsymbol{x}_{t}\cap X_{\mathcal{\mathcal{\mathcal{P}}}}^{\boldsymbol{\pi}}\neq\emptyset\right.\right]\\
 & \overset{_{^{\left(2\right)}}}{\geq}\mathbb{E}^{\boldsymbol{\pi}}\left[\gamma_{F}^{N_{t}}\cdot\mathcal{D}\left(\boldsymbol{x}_{t}\left[t+N_{t}:\right]\right)\left|\boldsymbol{x}_{t}\left[t+N_{t}\right]=x\right|\boldsymbol{x}_{t}\cap X_{\mathcal{\mathcal{\mathcal{P}}}}^{\boldsymbol{\pi}}\neq\emptyset\Bigr]\right.\\
 & \overset{_{^{\left(3\right)}}}{\geq}\mathbb{E}^{\boldsymbol{\pi}}\left[\gamma_{F}^{N_{t}}\Bigl|\right.\boldsymbol{x}_{t}\cap X_{\mathcal{\mathcal{\mathcal{P}}}}^{\boldsymbol{\pi}}\neq\emptyset\Bigr]\cdot U_{\min}^{\boldsymbol{\pi}}\left(\boldsymbol{x}_{t}\left[t+N_{t}\right]\right)\\
 & \overset{_{^{\left(4\right)}}}{\geq}\gamma_{F}^{\mathbb{E}^{\boldsymbol{\pi}}\left[N_{t}\left|\boldsymbol{x}_{t}\left[t\right]=x\right|\boldsymbol{x}_{t}\cap X_{\mathcal{\mathcal{\mathcal{P}}}}^{\boldsymbol{\pi}}\neq\emptyset\right]}\cdot U_{\min}^{\boldsymbol{\pi}}\left(x_{Acc}\right)\\
 & =\gamma_{F}^{n_{t}}\cdot U_{\min}^{\boldsymbol{\pi}}\left(x_{Acc}\right),
\end{aligned}
\label{eq:proof 2}
\end{equation}
where $x_{Acc}\in X_{\mathcal{\mathcal{\mathcal{P}}}}^{\boldsymbol{\pi}}$,
$U_{\min}^{\boldsymbol{\pi}}\left(x_{Acc}\right)=\min_{x\in X_{\mathcal{\mathcal{\mathcal{P}}}}^{\boldsymbol{\pi}}}U^{\boldsymbol{\pi}}\left(x\right)$,
and $n_{t}$ is a constant. By Lemma \ref{lemma:1}, one has $\underset{\gamma_{F}\shortrightarrow1^{-}}{\lim}U_{\min}^{\boldsymbol{\pi}}\left(x_{Acc}\right)=~1$.
In (\ref{eq:proof 2}), the first inequality \textbf{(1)} holds because visiting $X_{\mathcal{\mathcal{\mathcal{P}}}}^{\boldsymbol{\pi}}$
is one of the cases for $\diamondsuit F_{i}^{\mathcal{\mathcal{P}}}$ so that $\boldsymbol{x}_{t}\models\diamondsuit F_{i}^{\mathcal{\mathcal{P}}}$, e.g., $F_{i}^{\mathcal{\mathcal{P}}}$ can be placed outside of all BSCCs;
the second inequality \textbf{(2)} holds due to Lemma \ref{lemma:1}; the third
inequality \textbf{(3)} holds due to the Markov properties of (\ref{eq:DisctRetrn})
and (\ref{eq:ExpRetrn}); the fourth inequality \textbf{(4)} holds due to Jensen's
inequality. Based on (\ref{eq:proof 2}), the lower bound of (\ref{eq:proof 1})
is $U^{\boldsymbol{\pi}}\left(x\right)\geq\gamma_{F}^{n_{t}}\cdot U_{\min}^{\boldsymbol{\pi}}\left(x_{Acc}\right)\cdot\Pr{}^{\boldsymbol{\pi}}\left[x\models\diamondsuit F_{i}^{\mathcal{\mathcal{P}}}\right]$
from which one has 
\begin{equation}
\underset{\gamma_{F}\shortrightarrow1^{-}}{\lim}U^{\boldsymbol{\pi}}\left(x\right)\geq\gamma_{F}^{n_{t}}\cdot\Pr{}^{\boldsymbol{\pi}}\left[x\models\diamondsuit F_{i}^{\mathcal{\mathcal{P}}}\right].\label{eq:lower bound}
\end{equation}

Similarly, let $t+M_{t}$ denote the index that $\boldsymbol{x}_{t}$
first enters a BSCC that contains no accepting states. We have
\begin{equation}
\begin{array}{c}
\mathbb{E}^{\boldsymbol{\pi}}\left[\mathcal{D}\left(\boldsymbol{x}_{t}\right)\left|\boldsymbol{x}_{t}\mid\neq\diamondsuit F_{i}^{\mathcal{\mathcal{P}}}\right.\right]\overset{_{^{\left(1\right)}}}{\leq}\mathbb{E}^{\pi}\left[1-r_{F}^{M_{t}}\left|\boldsymbol{x}_{t}\mid\neq\diamondsuit F_{i}^{\mathcal{\mathcal{P}}}\right.\right]\\
\overset{_{^{\left(2\right)}}}{\leq}1-r_{F}^{\mathbb{E}^{\boldsymbol{\pi}}\left[M_{t}\left|\boldsymbol{x}_{t}\left[t\right]=x,\right.\boldsymbol{x}_{t}\mid\neq\diamondsuit F^{\mathcal{P}}\right]}=1-r_{F}^{m_{t}}
\end{array}\label{eq:proof 3}
\end{equation}
where $m_{t}$ is a constant and (\ref{eq:proof 3}) holds due to
Lemma \ref{lemma:1} and Markov properties.

Hence, the upper bound of (\ref{eq:proof 1}) is obtained as 
\begin{equation}
\underset{\gamma_{F}\shortrightarrow1^{-}}{\lim}U^{\boldsymbol{\pi}}\left(x\right)\leq\Pr{}^{\boldsymbol{\pi}}\left[x\models\diamondsuit F_{i}^{\mathcal{\mathcal{P}}}\right]+\left(1-r_{F}^{m_{t}}\right)\Pr{}^{\boldsymbol{\pi}}\left[x\mid\neq\diamondsuit F_{i}^{\mathcal{\mathcal{P}}}\right].\label{eq:upper bound}
\end{equation}
By (\ref{eq:lower bound}) and (\ref{eq:upper bound}), we can conclude
\[
\begin{array}{c}
\gamma_{F}^{n_{t}}\cdot\Pr{}^{\boldsymbol{\pi}}\left[x\models\diamondsuit F_{i}^{\mathcal{\mathcal{P}}}\right]\leq\underset{\gamma_{F}\shortrightarrow1^{-}}{\lim}U^{\boldsymbol{\pi}}\left(x\right)\\
\leq\Pr{}^{\boldsymbol{\pi}}\left[x\models\diamondsuit F_{i}^{\mathcal{\mathcal{P}}}\right]+\left(1-r_{F}^{m_{t}}\right)\cdot\Pr{}^{\boldsymbol{\pi}}\left[x\mid\neq\diamondsuit F_{i}^{\mathcal{\mathcal{P}}}\right]
\end{array}
\]
According to $\underset{\gamma_{F}\shortrightarrow1^{-}}{\lim}r_{F}\left(\gamma_{F}\right)=1$
in the reward function, (\ref{eq:reachability}) can be concluded. 
\end{IEEEproof}
When the condition $\gamma_{F}\shortrightarrow1^{-}$ holds, \cite{Bozkurt2020}
proves the expected return as the probability of satisfying the accepting
condition of LDBA. Different from \cite{Bozkurt2020}, Theorem \ref{lemma:probability}
only states that the expected return indicates the probability of
visiting an accepting set, rather than showing the probability of
satisfying the acceptance condition of E-LDGBA. Nevertheless, we will
show in the following section how Theorem \ref{lemma:probability}
can be leveraged to solve Problem \ref{Prob:2}. 
\begin{thm}
	\label{thm2}Consider a PL-MDP $\mathcal{M}$ and an E-LDGBA $\mathcal{\overline{A}}_{\phi}$
	corresponding to an LTL formula $\phi$. Based on
		assumption \ref{Assumption1}, there exists a discount factor $\underline{\gamma}$
		and any optimization method for (\ref{eq:ExpRetrn}) with $\gamma_{F}>\underline{\gamma}$
		and $r_{F}>\underline{\gamma}$ to to obtain a policy $\bar{\boldsymbol{\pi}}$
		, then the induced run $r_{\mathcal{\mathcal{\mathcal{\mathcal{\mathcal{\mathcal{P}}}}}}}^{\bar{\boldsymbol{\pi}}}$ satisfies the accpeting condition
		of the corresponding $\mathcal{\mathcal{\mathcal{\mathcal{P}}}}$
		(Def. \ref{def:P-MDP}). 
\end{thm}
\begin{IEEEproof}
	For any policy $\boldsymbol{\pi}$, $MC_{\mathcal{\mathcal{{\mathcal{P}}}}}^{\boldsymbol{\pi}}=\ensuremath{\mathcal{T}_{\boldsymbol{\pi}}}\sqcup\ensuremath{\mathcal{R}_{{\boldsymbol{\pi}}}^{1}\sqcup\ensuremath{\mathcal{R}_{{\boldsymbol{\pi}}}^{2}}\ldots\ensuremath{\mathcal{R}_{{\boldsymbol{\pi}}}^{n_{R}}}}.$
	Let $\boldsymbol{U}_{\boldsymbol{\pi}}=\left[\begin{array}{ccc}
	U^{\boldsymbol{\pi}}\left(x_{0}\right) & U^{\boldsymbol{\pi}}\left(x_{1}\right) & \ldots\end{array}\right]^{T}\in\mathbb{R}^{\left|X\right|}$ denote the stacked expected return under policy $\boldsymbol{\pi}$, which can
	be reorganized as 
	\begin{equation}
		\begin{aligned}\left[\begin{array}{c}
				\boldsymbol{U}_{\boldsymbol{\pi}}^{tr}\\
				\boldsymbol{U}_{\boldsymbol{\pi}}^{rec}
			\end{array}\right]= & \stackrel[n=0]{\infty}{\sum}\left(\stackrel[j=0]{n-1}{\prod}\left[\begin{array}{cc}
				\boldsymbol{\boldsymbol{\gamma}}_{\boldsymbol{\pi}}^{\ensuremath{\mathcal{T}}} & \boldsymbol{\boldsymbol{\gamma}}_{\boldsymbol{\pi}}^{\ensuremath{tr}}\\
				\boldsymbol{0}_{\sum_{i=1}^{m}N_{i}\times r} & \boldsymbol{\boldsymbol{\gamma}}_{\boldsymbol{\pi}}^{rec}
			\end{array}\right]\right)\\
			& \cdot\left[\begin{array}{cc}
				\boldsymbol{P}_{\boldsymbol{\pi}}\left(\ensuremath{\mathcal{T}},\ensuremath{\mathcal{T}}\right) & \boldsymbol{P}_{\boldsymbol{\pi}}^{tr}\\
				\boldsymbol{0}_{\sum_{i=1}^{m}N_{i}\times r} & \boldsymbol{P}_{\boldsymbol{\pi}}\left(\mathcal{R},\mathcal{R}\right)
			\end{array}\right]^{n}\left[\begin{array}{c}
				\boldsymbol{R}_{\boldsymbol{\pi}}^{tr}\\
				\boldsymbol{R}_{\boldsymbol{\pi}}^{rec}
			\end{array}\right],
		\end{aligned}
		\label{eq: utility_function}
	\end{equation}
	where $\boldsymbol{U}_{\boldsymbol{\pi}}^{tr}$ and $\boldsymbol{U}_{\boldsymbol{\pi}}^{rec}$
	are the expected return of states in transient and recurrent classes
	under policy $\boldsymbol{\pi}$, respectively. In (\ref{eq: utility_function}),
	$\boldsymbol{P}_{\pi}\left(\ensuremath{\mathcal{T}},\ensuremath{\mathcal{T}}\right)\in\mathbb{R}^{r\times r}$
	is the probability transition matrix between states in $\ensuremath{\mathcal{T}_{\boldsymbol{\pi}}}$,
	and $\boldsymbol{P}_{\boldsymbol{\pi}}^{tr}=\left[P_{\boldsymbol{\pi}}^{tr_{1}}\ldots P_{\boldsymbol{\pi}}^{tr_{m}}\right]\in\mathbb{R}^{r\times\sum_{i=1}^{m}N_{i}}$
	is the probability transition matrix where $P_{\boldsymbol{\pi}}^{tr_{i}}\mathbb{\in R}^{r\times N_{i}}$
	represents the transition probability from a transient state in $\ensuremath{\mathcal{T}_{\boldsymbol{\pi}}}$
	to a state of $\mathcal{R}_{\boldsymbol{\pi}}^{i}$. The $\boldsymbol{P}_{\boldsymbol{\pi}}\left(\mathcal{R},\mathcal{R}\right)$
	is a diagonal block matrix, where the $i$th block is a $N_{i}\times N_{i}$
	matrix containing the transition probabilities between states within $\mathcal{R}_{\boldsymbol{\pi}}^{i}$.
	Note that $\boldsymbol{P}_{\boldsymbol{\pi}}\left(\mathcal{R},\mathcal{R}\right)$
	is a stochastic matrix since each block matrix is a stochastic matrix
	\cite{Durrett1999}. Similarly, the rewards $\boldsymbol{\boldsymbol{R}}_{\pi}$
	can also be partitioned into $\boldsymbol{R}_{\boldsymbol{\pi}}^{tr}$
	and $\boldsymbol{R}_{\boldsymbol{\pi}}^{rec}$.
	
	The following proof is based on contradiction. Suppose there exists
	a policy $\boldsymbol{\pi}^{*}$ that optimizes the expected return, but not satisfy
	the accepting condition of $\mathcal{\mathcal{\mathcal{\mathcal{P}}}}$.
	Based on Lemma \ref{lemma:accepting set}, the following is true:
	$F_{k}^{\mathcal{\mathcal{\mathcal{\mathcal{\mathcal{P}}}}}}\subseteq\ensuremath{\mathcal{T}_{\boldsymbol{\pi}^{*}}},\forall k\in\left\{ 1,\ldots f\right\} $,
	where $\ensuremath{\mathcal{T}_{\boldsymbol{\boldsymbol{\pi}}^{*}}}$ denotes the
	transient class of Markov chain induced by $\boldsymbol{\pi}^{*}$ on $\mathcal{\mathcal{\mathcal{\mathcal{\mathcal{P}}}}}$.
	First, consider a state $x_{R}\in\mathcal{R}_{\boldsymbol{\pi}^{*}}^{j}$
	and let $\boldsymbol{P}_{\boldsymbol{\pi}^{*}}^{x_{R}R_{j}}$ denote
	a row vector of $\boldsymbol{P}_{\boldsymbol{\pi}^{*}}^{n}\left(\mathcal{R},\mathcal{R}\right)$
	that contains the transition probabilities from $x_{R}$ to the states
	in the same recurrent class $\mathcal{R}_{\boldsymbol{\pi}^{*}}^{j}$ after $n$
	steps. The expected return of $x_{R}$ under $\boldsymbol{\pi}^{*}$ is then obtained
	from (\ref{eq: utility_function}) as 
	\[
	U_{\boldsymbol{\pi}^{*}}^{rec}\left(x_{R}\right)=\stackrel[n=0]{\infty}{\sum}\gamma^{n}\left[\boldsymbol{0}_{k_{1}}^{T}\:\boldsymbol{P}_{\pi^{*}}^{x_{R}R_{j}}\:\boldsymbol{0}_{k_{2}}^{T}\right]\boldsymbol{R}_{\boldsymbol{\pi}^{*}}^{rec},
	\]
	where $k_{1}=\sum_{i=1}^{j-1}N_{i}$, $k_{2}=\sum_{i=j+1}^{n}N_{i}$.
	Since $\ensuremath{\mathcal{R}_{\boldsymbol{\pi}^{*}}^{j}}\cap F_{i}^{\mathcal{P}}=\emptyset,\forall i\in\left\{ 1,\ldots f\right\} $,
	by the designed reward function, all entries of $\boldsymbol{R}_{\boldsymbol{\pi}^{*}}^{rec}$
	are zero. We can conclude $U_{\boldsymbol{\pi}^{*}}^{rec}\left(x_{R}\right)=0$.
	To show contradiction, the following analysis will show that $U_{\bar{\boldsymbol{\pi}}}^{rec}\left(x_{R}\right)>U_{\boldsymbol{\pi}^{*}}^{rec}\left(x_{R}\right)$
	for any policy $\bar{\boldsymbol{\pi}}$ that satisfies the accepting condition
	of $\mathcal{\mathcal{P}}$. Thus, it's true that there exists $\mathcal{R}_{\bar{\boldsymbol{\pi}}}^{j}$
	such that $\mathcal{R}_{\bar{\boldsymbol{\pi}}}^{j}\cap F_{k}^{\mathcal{\mathcal{\mathcal{\mathcal{P}}}}}\neq\emptyset,\forall k\left\{ 1,\ldots f\right\} $.
	We use $\underline{\gamma}$ and $\overline{\gamma}$ to denote the
	lower and upper bound of $\gamma$.
	
	\textbf{Case 1:} If $x_{R}\in\mathcal{R}_{\bar{\boldsymbol{\pi}}}^{j}$,
	there exist states such that $x_{\varLambda}\in\mathcal{R}_{\bar{\boldsymbol{\pi}}}^{j}\cap F_{i}^{\mathcal{\mathcal{\mathcal{\mathcal{P}}}}}$.
	From Lemma \ref{lemma:accepting set}, the entries in $\boldsymbol{R}_{\bar{\boldsymbol{\pi}}}^{rec}$
	corresponding to the recurrent states in $\mathcal{R}_{\bar{\boldsymbol{\pi}}}^{j}$
	have non-negative rewards and at least there exist $f$ states in
	$\mathcal{R}_{\bar{\boldsymbol{\pi}}}^{j}$ from different accepting
	sets $F_{i}^{\mathcal{R}}$ with positive reward $r_{F}$. From (\ref{eq: utility_function}),
	$U_{\bar{\boldsymbol{\pi}}}^{rec}\left(x_{R}\right)$ can be lower
	bounded as 
	\[
	\begin{aligned}U_{\bar{\boldsymbol{\pi}}}^{rec}\left(x_{R}\right) & \geq\stackrel[n=0]{\infty}{\sum}\underline{\gamma}^{n}\left(P_{\bar{\boldsymbol{\pi}}}^{x_{R}x_{\varLambda}}r_{F}\right)\end{aligned}
	>0,
	\]
	where $P_{\bar{\boldsymbol{\pi}}}^{x_{R}x_{\varLambda}}$ is the transition
	probability from $x_{R}$ to $x_{\varLambda}$ in $n$ steps. We can
	conclude in this case $U_{\bar{\boldsymbol{\pi}}}^{rec}\left(x_{R}\right)>U_{\boldsymbol{\pi}^{*}}^{rec}\left(x_{R}\right)$.
	
	\textbf{Case 2:} If $x_{R}\in\mathcal{T}_{\bar{\boldsymbol{\pi}}}$,
	there are no states of any accepting set $F_{i}^{\mathcal{P}}$ in
	$\mathcal{T}_{\bar{\boldsymbol{\pi}}}$. As demonstrated in \cite{Durrett1999},
	for a transient state $x_{tr}\in\mathcal{T}_{\bar{\boldsymbol{\pi}}}$, there always
	exists an upper bound $\Delta<\infty$ such that $\stackrel[n=0]{\infty}{\sum}p^{n}\left(x_{tr},x_{tr}\right)<\Delta$,
	where $p^{n}\left(x_{tr},x_{tr}\right)$ denotes the probability of
	returning from a transient state $x_{T}$ to itself in $n$ time steps.
	In addition, for a recurrent state $x_{rec}$ of $\mathcal{R}_{\bar{\boldsymbol{\pi}}}^{j}$,
	it is always true that 
	\begin{equation}
		\stackrel[n=0]{\infty}{\sum}\gamma^{n}p^{n}\left(x_{rec},x_{rec}\right)>\frac{1}{1-\gamma^{\overline{n}}}\bar{p},\label{eq:case2_1}
	\end{equation}
	where there exists $\overline{n}$ such that $p^{\overline{n}}\left(x_{rec},x_{rec}\right)$
	is nonzero and can be lower bounded by $\bar{p}$ \cite{Durrett1999}.
	From (\ref{eq: utility_function}), one has 
	\begin{equation}
		\begin{aligned}\boldsymbol{U}_{\bar{\boldsymbol{\boldsymbol{\pi}}}}^{tr} & >\stackrel[n=0]{\infty}{\sum}\left(\stackrel[j=0]{n-1}{\prod}\boldsymbol{\boldsymbol{\gamma}}_{\boldsymbol{\pi}}^{tr}\right)\ldotp\boldsymbol{P}_{\bar{\boldsymbol{\pi}}}^{tr}\boldsymbol{P}_{\bar{\boldsymbol{\pi}}}^{n}\left(\mathcal{R},\mathcal{R}\right)\boldsymbol{R}_{\boldsymbol{\pi}}^{rec}\\
			& {\color{black}>\underline{\gamma}^{n}\ldotp\boldsymbol{P}_{\bar{\boldsymbol{\pi}}}^{tr}\boldsymbol{P}_{\bar{\boldsymbol{\pi}}}^{n}\left(\mathcal{R},\mathcal{R}\right)\boldsymbol{R}_{\boldsymbol{\pi}}^{rec}}.
		\end{aligned}
		\label{eq:case2_2}
	\end{equation}
	Let $\max\left(\cdot\right)$ and $\min\left(\cdot\right)$ represent
	the maximum and minimum entry of an input vector, respectively. The
	upper bound $\bar{m}=\left\{ \max\left(\overline{M}\right)\left|\overline{M}<\boldsymbol{P}_{\bar{\boldsymbol{\pi}}}^{tr}\boldsymbol{\bar{P}}\boldsymbol{R}_{\boldsymbol{\pi}}^{rec}\right.\right\} $
	and $\bar{m}\geq0$, where $\boldsymbol{\bar{P}}$ is a block matrix
	whose nonzero entries are derived similarly to $\bar{p}$ in (\ref{eq:case2_1}).
	The utility $U_{\bar{\pi}}^{tr}\left(x_{R}\right)$ can be lower bounded
	from (\ref{eq:case2_1}) and (\ref{eq:case2_2}) as 
	\begin{equation}
		U_{\bar{\boldsymbol{\pi}}}^{tr}\left(x_{R}\right)>\frac{1}{1-\underline{\gamma}^{n}}\bar{m}.\label{eq:case2_3}
	\end{equation}
	Since $U_{\boldsymbol{\pi}^{*}}^{rec}\left(x_{R}\right)=0$, the contradiction
	$U_{\bar{\boldsymbol{\pi}}}^{tr}\left(x_{R}\right)>0$ is achieved
	if $\frac{1}{1-\underline{\gamma}^{n}}\bar{m}$. Thus, there exist
	$0<\underline{\gamma}<1$ such that $\gamma_{F}>\underline{\gamma}$
	and $r_{F}>\underline{\gamma}$, which implies $U_{\bar{\boldsymbol{\pi}}}^{tr}\left(x_{R}\right)>\frac{1}{1-\underline{\gamma}^{n}}\bar{m}\geq0$.
	The procedure shows the contradiction of the assumption that $\boldsymbol{\pi}^{*}$
	that does not satisfy the acceptance condition of $\mathcal{\mathcal{\mathcal{\mathcal{\mathcal{P}}}}}$
	is optimal, and Theorem \ref{thm2} is proved. 
\end{IEEEproof}
Theorem \ref{thm2} proves that by selecting $\gamma_{F}>\underline{\gamma}$
and $r_{F}>\underline{\gamma}$, optimizing the expected return in
(\ref{eq:ExpRetrn}) can find a policy satisfying the given task $\phi$. 
\begin{thm}
	\label{thm:probability} Given a PL-MDP $\mathcal{M}$ and an E-LDGBA
	$\mathcal{\overline{A}}_{\phi}$, by selecting $\gamma_{F}\shortrightarrow1^{-}$,
	the optimal policy $\boldsymbol{\pi}^{*}$ that maximizes the expected
	return (\ref{eq:ExpRetrn}) of the corresponding EP-MDP also
	maximizes the probability of satisfying $\phi$, i.e., $\Pr^{\boldsymbol{\pi}^{*}}\left[x\models Acc_{\mathcal{P}}\right]=\Pr_{max}\left[x\models Acc_{\mathcal{P}}\right]$. 
\end{thm}
\begin{IEEEproof}
	Since $\gamma_{F}\shortrightarrow1^{-}$, we have $\gamma_{F}>\underline{\gamma}$
	and $r_{F}>\underline{\gamma}$ from Theorem \ref{thm2}. There exists
	an induced run $r_{\mathcal{\mathcal{\mathcal{\mathcal{\mathcal{\mathcal{P}}}}}}}^{\boldsymbol{\pi}^{*}}$
	satisfying the accepting condition of $\mathcal{\mathcal{\mathcal{\mathcal{P}}}}$.
	According to Lemma \ref{lemma:probability}, $\underset{\gamma_{F}\shortrightarrow1^{-}}{\lim}U^{\boldsymbol{\pi}^{*}}\left(x\right)$
	is exactly equal to the probability of visiting the accepting sets
	of an AMEC. Optimizing $\underset{\gamma_{F}\shortrightarrow1^{-}}{\lim}U^{\boldsymbol{\pi}^{*}}\left(x\right)$
	is equal to optimizing the probability of entering AMECs. 
\end{IEEEproof}

\subsection{Model-Free Reinforcement Learning\label{subsec:RL}}

Based on the Q-learning \cite{Watkins1992}, the agent updates its
Q-value from $x$ to $x'$ according to 
\begin{equation}
	\begin{aligned}Q\left(x,u^{\mathcal{\mathcal{\mathcal{\mathcal{\mathcal{\mathcal{P}}}}}}}\right)\text{\ensuremath{\leftarrow}} & \left(1-\alpha\right)Q\left(x,u^{\mathcal{\mathcal{\mathcal{\mathcal{\mathcal{P}}}}}}\right)\\
		& +\alpha\left[R\left(x\right)+\gamma\left(x\right)\cdot\underset{\overline{u}^{\mathcal{\mathcal{\mathcal{\mathcal{\mathcal{\mathcal{P}}}}}}}\in U^{\mathcal{\mathcal{\mathcal{\mathcal{\mathcal{\mathcal{P}}}}}}}}{\max}Q\left(x',\overline{u}^{\mathcal{\mathcal{\mathcal{\mathcal{\mathcal{\mathcal{P}}}}}}}\right)\right],
	\end{aligned}
	\label{eq:Q}
\end{equation}
where $Q\left(x,u^{\mathcal{\mathcal{P}}}\right)$ is the Q-value
of the state-action pair $\left(x,u^{\mathcal{\mathcal{P}}}\right)$,
$0<\alpha\leq1$ is the learning rate, $0\leq\gamma\left(x\right)\leq1$
is the discount function as defined in section \ref{subsec:RL-reward}.
With the standard learning rate and discount factor, Q-value will
converge to a unique limit $Q^{*}$ as in \cite{Watkins1992}. Therefore,
the optimal expected utility and policy can be obtained as $U_{\boldsymbol{\pi}}^{*}\left(x\right)=\underset{u^{\mathcal{\mathcal{P}}}\in U^{\mathcal{\mathcal{P}}}\left(x\right)}{\max}Q^{*}\left(x,u^{\mathcal{\mathcal{P}}}\right)$
and $\boldsymbol{\pi}^{*}\left(x\right)=\underset{u^{\mathcal{\mathcal{P}}}\in U^{\mathcal{\mathcal{P}}}\left(x\right)}{\arg\max}Q^{*}\left(x,u^{\mathcal{\mathcal{P}}}\right)$.

\begin{algorithm}
	\caption{\label{Alg2} RL based motion planning under LTL}
	
	\scriptsize
	
	\singlespacing
	
	\begin{algorithmic}[1]
		
		\Procedure {Input: } {$\mathcal{M}$ , $\phi$, $\varLambda$}
		
		{Output: } { optimal policy $\boldsymbol{\pi}^{*}$ }
		
		{Initialization: } { Set $episode=0$ , $iteration=0$ and $\tau$
			(maximum allowed learning steps) }
		
		\State set $r_{F}=0.99$ and $\gamma_{F}=0.9999$ to determine $R\left(x\right)$
		and $\gamma\left(x\right)$
		
		\For { all $x\in X$ }
		
		\State $U\left(x\right)=0$ and $Q\left(x,u^{\mathcal{\mathcal{\mathcal{\mathcal{\mathcal{P}}}}}}\right)=0$,
		 $\forall u^{\mathcal{\mathcal{\mathcal{\mathcal{P}}}}}\in U^{\mathcal{P}}\left(x\right)$
		
		\State $Count\left(x,u^{\mathcal{\mathcal{\mathcal{\mathcal{\mathcal{P}}}}}}\right)=0$, $\forall u^{\mathcal{\mathcal{\mathcal{\mathcal{P}}}}}\in U^{\mathcal{P}}\left(x\right)$
		
		\EndFor
		
		\State $x_{curr}=x_{0}$;
		
		\While { $\boldsymbol{U}$ are not converged }
		
		\State $episode++$;
		
		\State $\epsilon=1/episode$;
		
		\While {$iteration<\tau$ }
		
		\State $iteration++$
		
		\State Select $u^{\mathcal{P}}_{curr}$ based on epsilon-greedy selection
		
		\State $\text{ }$
		
		\State Execute $u^{\mathcal{P}}_{curr}$ and observer $x_{next}$, $R\left(x_{curr}\right)$,$\gamma\left(x_{curr}\right)$
		
		\State $r\text{\ensuremath{\leftarrow}}R\left(x_{curr}\right)$ and
		$\gamma\text{\ensuremath{\leftarrow}}\gamma\left(x_{curr}\right)$
		
		\State $Count\left(x_{curr},u^{\mathcal{P}}_{curr}\right)++$
		
		\State $\alpha=1/Count\left(x_{curr},u^{\mathcal{P}}_{curr}\right)$
		
		\State $Q\left(x_{curr},u^{\mathcal{P}}_{curr}\right)\text{\ensuremath{\leftarrow}}\left(1-\alpha\right)Q\left(x_{curr},u^{\mathcal{P}}_{curr}\right)+\newline\hspace*{4.8em}\alpha\left[r+\gamma\cdot\underset{u^{\mathcal{\mathcal{\mathcal{\mathcal{\mathcal{P}}}}}}\in U^{\mathcal{\mathcal{\mathcal{P}}}}}{\max}Q\left(x_{next},u^{\mathcal{\mathcal{\mathcal{\mathcal{\mathcal{P}}}}}}\right)\right]$
		
		\State $x_{curr}=x_{next}$
		
		\EndWhile
		
		\EndWhile
		
		\For { all $x\in X$ }
		
		\State $\boldsymbol{\pi}^{*}\left(x\right)=\underset{u^{\mathcal{\mathcal{\mathcal{\mathcal{\mathcal{P}}}}}}\in U^{\mathcal{\mathcal{\mathcal{P}}}}}{\max}Q\left(x,u^{\mathcal{\mathcal{\mathcal{\mathcal{\mathcal{P}}}}}}\right)$
		
		\EndFor
		
		\EndProcedure
		
	\end{algorithmic}
\end{algorithm}

\footnote{Any other model-free RL algorithm can also be adopted with Alg.\ref{Alg2}}The
learning strategy is outlined in Alg. \ref{Alg2}. In
	line 2, the dependent and discount values are selected based on section
\ref{subsec:RL-reward}. 

\section{Case Studies\label{sec:Case}}

The developed RL-based control synthesis is implemented in Python.
Owl \cite{Kretinsky2018} is used to convert LTL specifications to
LDGBA. We implement the Alg. \ref{Alg2} to validate the
effectiveness of our approach, and we first carry out simulations over
grid environments and then validate the approach in a more realistic
office scenario with TurtleBot3 robot, which consider the uncertainties of both motion and environment. The software can be find in our Github repository\footnote{\url{https://github.com/mingyucai/E-LDGBA_RL}}.

\subsection{Simulation Results\label{subsec:Simulation-Results}}

\begin{figure}
	\centering{}\includegraphics[scale=0.50]{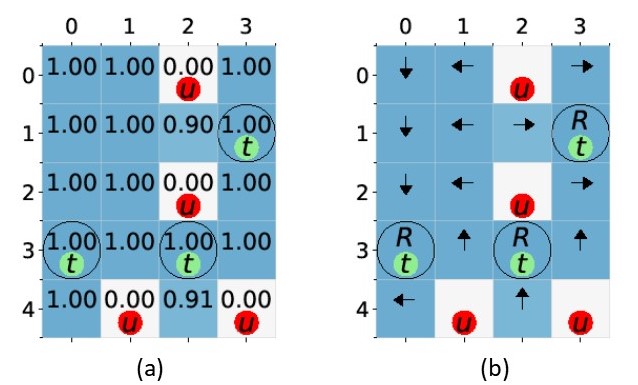}\caption{\label{fig:case_study1} (a) The estimated maximal satisfaction probability.,
		where the targets and unsafe areas are denoted by $\mathtt{t}$ and
		$\mathtt{u}$, respectively. (b) The optimal policy of satisfying
		$\varphi_{case1}$, where ``$R$'' is an action primitive means
		the robot remains at its current cell.}
\end{figure}

\begin{figure}
	\centering{}\includegraphics[scale=0.35]{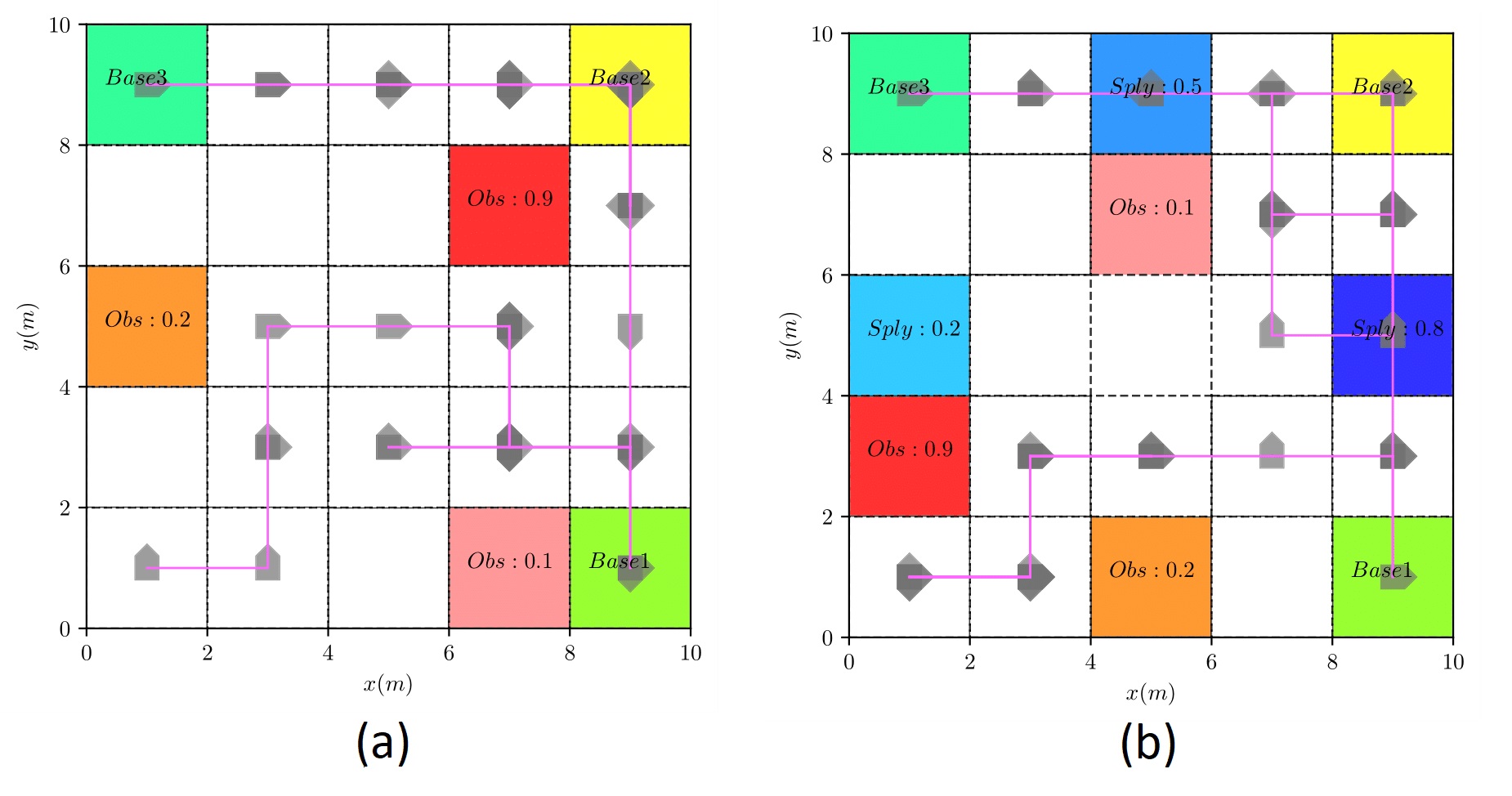}\caption{\label{fig:case_study2} Simulated trajectories of 25 time steps under
		the corresponding optimal policies.} 
\end{figure}

Consider a mobile robot following the unicycle model, i.e. $\dot{x}=v\sin\left(\theta\right)$,
$\dot{y}=v\cos\left(\theta\right)$, and $\dot{\theta}=\omega$, where
$x,y,\theta$ indicate the robot positions and orientation. The linear
and angular velocities are the control inputs, i.e., $u=\left(v,\omega\right)$.
The workspace is shown in Fig. \ref{fig:case_study1} and Fig. \ref{fig:case_study2}.
To model motion uncertainties, we assume the action primitives can
not always be successfully executed. For instance, action primitives
``$N,S,E,W$'' mean the robot can successfully move towards north,
south, east and west (four possible orientations) to adjacent cells
with probability $0.9$, respectively, and fails by moving sideways
with probability $0.1$. Action primitive ``$R$'' means the robot
remains at its current cell.

\textbf{(1) Maximum Satisfaction Probability}: In this case, the objective
is to verify that the generated policy satisfies the LTL specification
with a maximum probability. The package Csrl in \cite{Bozkurt2020}
is used. The LTL specification is 
\begin{equation}
\varphi_{case1}=\lozenge\oblong\mathtt{t}\land\oblong\lnot\mathtt{u},\label{eq:case_probability}
\end{equation}
which requires the robot to eventually arrive at one of the targets
$\mathtt{t}$ while avoiding unsafe areas $\mathtt{u}$. Each episode terminates after $\tau=100$ steps. Fig.
\ref{fig:case_study1} (a) shows the estimated maximum probability
of satisfying $\varphi_{case1}$ starting from each state. Note that
the maximum satisfaction probability starting from $(2,1)$ is $0.9$,
since the robot can move sideways with probability $0.1$ due to motion
uncertainties. Suppose the robot starts from $(0,0)$, Fig. \ref{fig:case_study1}
(b) shows the generated optimal policy at each state, and the robot
will complete $\varphi_{case1}$ with probability one by eventually
visiting either $(0,3)$ or $(2,3)$. After arriving at the destination,
the robot will select ``$R$'' to stay at the target.

We then verify more complex LTL specifications over the infinite horizon.
As shown in Fig. \ref{fig:case_study2}, the cells are marked with
different colors to represent different areas of interest, e.g., $\mathtt{Base}1,\mathtt{Base}2,\mathtt{Base}3,\mathtt{Obs},\mathtt{Sply}$,
where $\mathtt{Obs}$ and $\mathtt{Sply}$ are shorthands for obstacle
and supply, respectively. To model the environment uncertainties, the
number associated with a cell represents the likelihood that the corresponding
property appears at that cell. For example, $\mathtt{Obs}:0.1$ indicates
this cell is occupied by the obstacles with probability 0.1. In Fig.
\ref{fig:case_study2} (a), we first consider a case that user-specified
tasks can all be successfully executed. The desired surveillance task
to be performed is formulated as 
\[
\varphi_{case2}=\left(\oblong\lozenge\mathtt{Base}1\right)\land\left(\oblong\lozenge\mathtt{Base}2\right)\land\left(\oblong\lozenge\mathtt{Base}3\right)\land\oblong\lnot\mathtt{Obs},
\]
which requires the mobile robot to visit all base stations infinitely often 
while avoiding the obstacles. In
this case, each episode terminates after $\tau=100$ steps. The generated
optimal trajectory is shown in Fig. \ref{fig:case_study2} (a), which
indicates $\varphi_{case1}$ is completed. We then validate our approach
with more complex task specifications 
\[
\varphi_{case3}=\varphi_{case1}\land\oblong\left(\mathtt{Sply}\rightarrow\varbigcirc\left(\left(\lnot\mathtt{Sply}\right)\cup\mathtt{\varphi_{one1}}\right)\right),
\]
where $\varphi_{one1}=\mathtt{Base}1\lor\mathtt{Base}2\lor\mathtt{Base}3$.
$\varphi_{case3}$ requires the robot to visit the supply station
and then go to one of the base stations while avoiding obstacles and
requiring all base stations to be visited. In this case, each episode terminates after $\tau=800$
steps. The generated optimal trajectory is shown in Fig. \ref{fig:case_study2}
(b).

\begin{figure}
	\centering{}\includegraphics[scale=0.20]{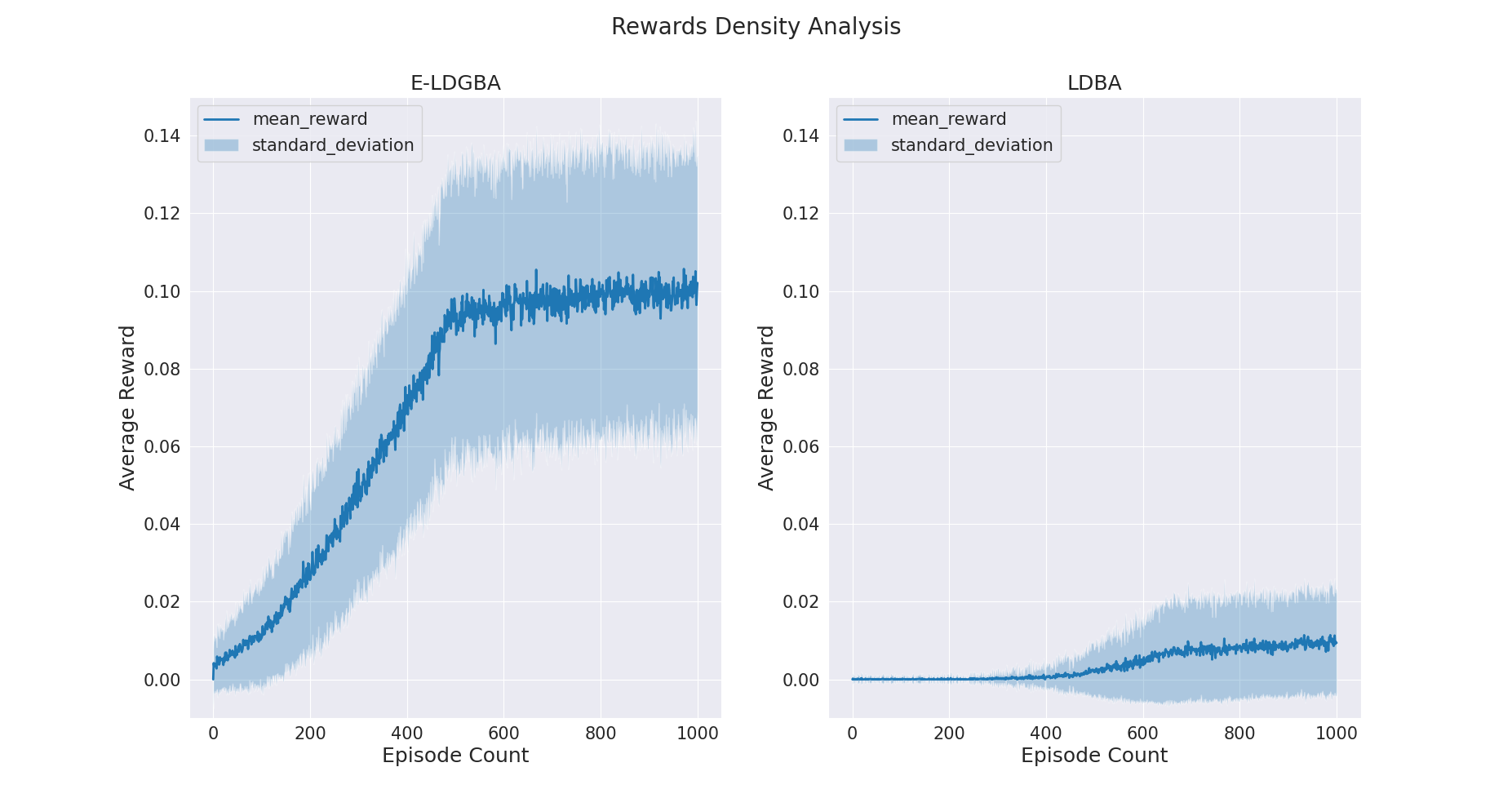}\caption{\label{fig:Reward_Density_Analysis}
		The mean and standard deviation of rewards via E-LDGBA (left) and LDBA (right)
		respectively}
\end{figure}

\textbf{(2) Reward Density}: Since LDBA can be considered as a special
case of E-LDGBA with only one accepting set, LDBA is compared with
E-LDGBA in this case. To show the benefits of applying E-LDGBA over
LDBA in overcoming the issues of spare rewards, we perform 100 learning
iterations for 1000 episodes and compare the reward collection in
the training process. The RL-based policy synthesis is carried out
for $\varphi_{case2}$. Fig. \ref{fig:Reward_Density_Analysis} shows
the mean and standard deviations of collected rewards using E-LDGBA
and LDBA, respectively. It can be seen that E-LDGBA based method converges
faster since the accepting set is tracked every time when the robot visits
one of the base stations. In contrast, LDBA only records the time when all
base states have been visited. Hence, the sparse reward issue is relaxed
in our method. 

\textbf{(3) Scalability}: To show the computational complexity, the
RL-based policy synthesis is also performed for $\varphi_{case2}$
over workspaces of various sizes (each grid is further partitioned).
The simulation results are listed in Table \ref{tab:case1}, which
consists of the number of MDP states, the number of relaxed product
MDP states. The steps in Table \ref{tab:case1} indicate the time
used to converge to an optimal satisfaction planing when applying
reinforcement learning. It is also verified that the given task $\varphi_{case2}$
can be successfully carried out in larger workspaces. 
\begin{table}
	\caption{\label{tab:case1}Simulation results of large scale workspaces}
	\centering{}%
	\begin{tabular}{c|cccc}
		\hline 
		Workspace & MDP & $\mathcal{R}$ & Episode & \tabularnewline
		size{[}cell{]} & States & States & Steps & \tabularnewline
		$15\times15$ & 225 & 450 & 800 & \tabularnewline
		$25\times25$ & 625 & 1250 & 2000 & \tabularnewline
		$40\times40$ & 1600 & 3200 & 5000 & \tabularnewline
		\hline 
	\end{tabular}
\end{table}

\subsection{Experimental Results\label{subsec:Experimental-Results}}

\begin{figure}
	\centering{}\includegraphics[scale=0.45]{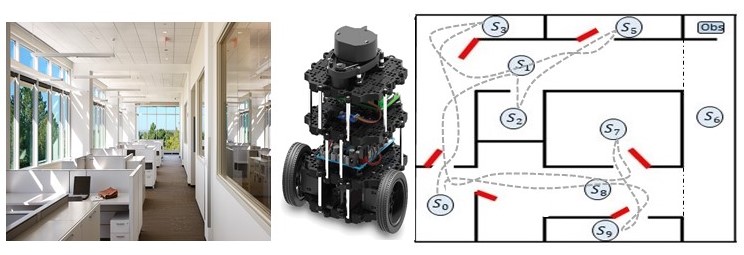}\caption{\label{fig:case_study3} \textcolor{black}{The mock-up office scenario
			with the }TurtleBot3\textcolor{black}{{} robot.}}
\end{figure}

Consider an office environment constructed in ROS Gazebo as shown
in Fig.\ref{fig:case_study3}, which consists of $7$ rooms denoted
by $S_{0},S_{2},S_{3},S_{5},S_{7},S_{9},\mathtt{Obs}$ and $5$ corridors
denoted by $S_{1},S_{6},S_{8}$. The two black dash lines are the
dividing lines for corridors $S_{1},S_{6}$ and $S_{6},S_{8}$, separately.
Starting from room $S_{0}$, the TurtleBot3 can follow a collision-free
path from the center of one region to another without crossing other
regions using obstacle-avoidance navigation. To model motion uncertainties,
it is assumed that the robot can successfully follow its navigation
controller moving from corridors to a desired room with probability
$0.9$ and fail by moving to the adjacent room with probability $0.1$. In addition, 
the robot can successfully moving between corridors with probability
$1.0$. To model environment uncertainties, we set the door of each room keeps open with probability
$0.9$ and close with probability $0.1$. The service to be performed
by TurtleBot3 is expressed as 
\begin{equation}
\varphi_{case4}=\varphi_{all}\land\oblong\lnot\mathtt{Obs},\label{eq:case4}
\end{equation}
where $\varphi_{all}=\oblong\lozenge S_{2}\land\oblong\lozenge S_{3}\land\oblong\lozenge S_{5}\land\oblong\lozenge S_{9}\land\oblong\lozenge S_{10}$.
In (\ref{eq:case4}), $\varphi_{all}$ requires the robot to always
service all rooms (e.g. pick trash) and return to $S_{0}$ (e.g. release
trash), while avoiding $\mathtt{Obs}$. The optimal policy for the case is generated that each episode
terminates after $\tau=150$ steps. The generated satisfying trajectories
(without collision) marked as gray bold dash line are shown in Fig.
\ref{fig:case_study3}. To maximize the satisfaction probability,
it is observed that the optimal policy avoids the corridor $S_{6}$,
since there is a non-zero probability of entering $\mathtt{Obs}$
from $S_{6}$ under any policies due to the uncertainties, resulting in
the violation of $\varphi_{case3}$. 

\section{Conclusion }
In this paper, the LTL specifications are translated to E-LDGBA to
apply the deterministic policy, and a model-free learning-based algorithm is developed
to synthesize control policies that maximize the satisfaction of LTL
specifications. Future research will focus on deep RL to address continuous
state and action spaces.

\bibliographystyle{IEEEtran}
\bibliography{references}

\end{document}